%% file: main.tex
\documentclass[acmsmall]{acmart}

\acmArticleSeq{0}
\acmSubmissionID{0}

\usepackage{array}
\usepackage{graphicx}
\usepackage{clrscode}
\usepackage{balance}
\usepackage{subfigure}
\usepackage{multirow}
\usepackage{booktabs}
\usepackage{adjustbox}
\usepackage[justification=justified,skip=5pt]{caption}
\usepackage{float}
\usepackage{color}
\usepackage{soul}
\usepackage{mdframed}
\usepackage{amsopn}
\usepackage{mathrsfs}
\usepackage{mathtools}
\usepackage{amsmath}
\usepackage{arydshln}
\usepackage{hyperref}
\usepackage{multicol}
\usepackage{blkarray}
\usepackage{enumerate}
\usepackage{courier}
\usepackage{rotating}
\usepackage{booktabs}
\usepackage{diagbox}
\usepackage{fancybox}
\usepackage{minibox}
\usepackage{cases}
\usepackage{bm}
\usepackage{enumitem}
\usepackage{xcolor}
\usepackage[lined, ruled, commentsnumbered, linesnumbered]{algorithm2e} %
\usepackage{enumitem}
\usepackage{tikz}
\usepackage{csquotes}

\setlist[itemize]{leftmargin=*}

\AtBeginDocument{%
  \providecommand\BibTeX{{%
    \normalfont B\kern-0.5em{\scshape i\kern-0.25em b}\kern-0.8em\TeX}}}

\setcopyright{rightsretained}
\copyrightyear{2024}
\acmYear{2024}

\acmJournal{TORS}
\acmVolume{1}
\acmNumber{1}
\acmArticle{1}
\acmMonth{1}

\makeatletter
\def\@tocline#1#2#3#4#5#6#7{\relax
  \ifnum #1>\c@tocdepth 
  \else
    \par \addpenalty\@secpenalty\addvspace{#2}%
    \begingroup \hyphenpenalty\@M
    \@ifempty{#4}{%
      \@tempdima\csname r@tocindent\number#1\endcsname\relax
    }{%
      \@tempdima#4\relax
    }%
    \parindent\z@ \leftskip#3\relax \advance\leftskip\@tempdima\relax
    \rightskip\@pnumwidth plus4em \parfillskip-\@pnumwidth
    #5\leavevmode\hskip-\@tempdima
      \ifcase #1
       \or\or \hskip 1em \or \hskip 2em \else \hskip 3em \fi%
      #6\nobreak\relax
    \hfill\hbox to\@pnumwidth{\@tocpagenum{#7}}\par
    \nobreak
    \endgroup
  \fi}
\makeatother

\begin{document}

\setcounter{secnumdepth}{2}
\setcounter{tocdepth}{2}
\title{A Survey on Trustworthy Recommender Systems}


\author{Yingqiang Ge}
\affiliation{%
  \institution{Department of Computer Science, Rutgers University, USA}
}
\email{yingqiang.ge@rutgers.edu}

\author{Shuchang Liu}
\affiliation{%
  \institution{Department of Computer Science, Rutgers University, USA}
}
\email{sl1471@scarletmail.rutgers.edu}

\author{Zuohui Fu}
\affiliation{%
  \institution{Department of Computer Science, Rutgers University, USA}
}
\email{zuofui.fu@rutgers.edu}

\author{Juntao Tan}
\affiliation{%
  \institution{Department of Computer Science, Rutgers University, USA}
}
\email{juntao.tan@rutgers.edu}

\author{Zelong Li}
\affiliation{%
  \institution{Department of Computer Science, Rutgers University, USA}
}
\email{zelong.li@rutgers.edu}

\author{Shuyuan Xu}
\affiliation{%
  \institution{Department of Computer Science, Rutgers University, USA}
}
\email{shuyuan.xu@rutgers.edu}

\author{Yunqi Li}
\affiliation{%
  \institution{Department of Computer Science, Rutgers University, USA}
}
\email{yunqi.li@rutgers.edu}


\author{Yikun Xian}
\affiliation{%
  \institution{Department of Computer Science, Rutgers University, USA}
}
\email{yx150@rutgers.edu}

\author{Yongfeng Zhang}
\affiliation{%
  \institution{Department of Computer Science, Rutgers University, USA}
}
\email{yongfeng.zhang@rutgers.edu}

\renewcommand{\shortauthors}{Y. Ge, S. Liu, Z. Fu, J. Tan, Z. Li, S. Xu, Y. Li, Y. Xian, Y. Zhang}


\begin{abstract}
Recommender systems (RS), serving at the forefront of Human-centered AI, are widely deployed in almost every corner of the web and facilitate the human decision-making process. However, despite their enormous capabilities and potential, RS may also lead to undesired effects on users, items, producers, platforms, or even the society at large, such as compromised user trust due to non-transparency, unfair treatment of different consumers, or producers, privacy concerns due to extensive use of user's private data for personalization, just to name a few. All of these create an urgent need for \textit{Trustworthy Recommender Systems (TRS)} so as to mitigate or avoid such adverse impacts and risks. In this survey, we will introduce techniques related to trustworthy recommendation, including but not limited to explainable recommendation, fairness in recommendation, privacy-aware recommendation, robustness in recommendation, user-controllable recommendation, as well as the relationship between these different perspectives in terms of trustworthy recommendation. Through this survey, we hope to deliver readers with a comprehensive view of the research area and raise attention to the community about the importance, existing research achievements, and future research directions on trustworthy recommendation.

\end{abstract}

\maketitle

\newpage
\tableofcontents
\newpage

\section{Introduction}
Recommender systems (RS)---which are extensively deployed in various systems such as e-commerce, social networks, search engines, news portals, hiring platforms, intelligent assistants, smart home and smart city services, as well as healthcare and financial applications---have been acknowledged for their capacity to deliver high-quality services that bridge the gap between users and items
by delivering tailored content for each individual.
Recommender systems not only help users to find relevant information more efficiently, but also directly influence the human decision-making process by providing relevant suggestions or even shaping users' worldviews by exposing users to the selected content. 

However, for every plus there is a minus; RS may offer both promise and perils. There are growing concerns that the irresponsible use of recommendation techniques may bring undesired effects and untrustworthy issues, such as compromised user trust due to non-transparency, unfair treatment of different users, producers, or platforms, privacy concerns due to extensive use of user's private data for personalization, {unsatisfied user experience due to the lack of user controllability} --- the list just continues to expand. These vulnerabilities significantly limit the development and deployment of recommendation algorithms and may even lead to severe economic and social issues. As a result, only considering recommendation accuracy is not enough when developing modern recommendation systems. We also need to make sure that the models are fair, have not been tampered with, will not fall apart in different conditions, and can be understood by humans. Moreover, the design and development process of RS also needs to be transparent and inclusive. All of these considerations beyond accuracy that make recommender systems safe, responsible, and worthy of our trust are related to trustworthy recommender systems research. In light of the growing need, the purpose of this survey is to provide a comprehensive overview of recent research efforts surrounding Trustworthy Recommender System (TRS).
Since recommender system is an important direction of Human-centered AI research that directly involves humans in the loop, Trustworthy Recommender System (TRS) has been leading the research of Trustworthy Artificial Intelligence (TAI) over the past years on various perspectives, such as the definition, method and evaluation of trustworthiness on explainability, fairness, robustness, and privacy, as well as how humans interact with trustworthy AI systems.

Semantically, the concept of trustworthiness relates to the ability to be trustworthy, dependable, reliable, responsible, faithful, honorable, creditworthy, and so on\footnote{https://www.merriam-webster.com/thesaurus/trustworthy}. 
The trustworthiness of a recommender system is based on the fundamental concept of ``trust'', which is what makes the system ``worthy'' of reliance. Trust can be defined as the willingness of one party (the trustor) to become vulnerable to another party (the trustee) on the presumption that the trustee will act in ways that benefit the trustor. \footnote{https://en.wikipedia.org/wiki/Trust\_(social\_science)} 
In the context of trustworthy recommender systems, the trustee is the recommender system itself, while the trustor may be its owners, consumers, producers, or even society as a whole. 

Due to its importance and necessity, there have been numerous discussions and debates over the  connotations of Trustworthy Artificial Intelligence (TAI).
In particular, \citeauthor{10.1145/3351095.3372834} \cite{10.1145/3351095.3372834} study trust in AI and summarized the attributes of TAI as Ability, Benevolence, Integrity, and Predictability;
\citeauthor{varshney2022trustworthy} \cite{varshney2022trustworthy} believe that a trustworthy machine learning system is one that has sufficient Basic Performance, Reliability, Human Interaction, and Selflessness; \citeauthor{liu2021trustworthy} \cite{liu2021trustworthy} consider TAI as a threat or risk-free program and focus on six dimensions in achieving trustworthiness: Safety \& Robustness, Nondiscrimination \& Fairness, Explainability, Privacy, Accountability \& Auditability, and Environmental Well-being. Moreover, in the year of 2019, European Union (EU) proposed the Ethics Guidelines for Trustworthy AI\footnote{\url{https://digital-strategy.ec.europa.eu/en/library/ethics-guidelines-trustworthy-ai}}, requiring that an AI system should meet four ethical principles: respect for human autonomy, prevention of harm, explicability, and fairness.
Although the existing literature explores the space of trustworthiness from various perspectives, several key aspects that have received the most recognition and consensus are \textit{explainability, fairness, privacy, robustness, and controllability}, which we believe are also the key components of Trustworthy Recommender Systems (TRS). Specifically, from a utility and economic perspective, an RS should provide accurate and robust recommendation results. Moreover, from a legal perspective, it should also protect users' privacy and offer explanations of its decision-making process. Finally, from an ethical perspective, it should mitigate the bias and unfairness issues and provide users with the ability to control the system so as to foster better understanding and trust in the system.


\begin{figure}[t]
  \centering
  \includegraphics[width=0.65\textwidth]{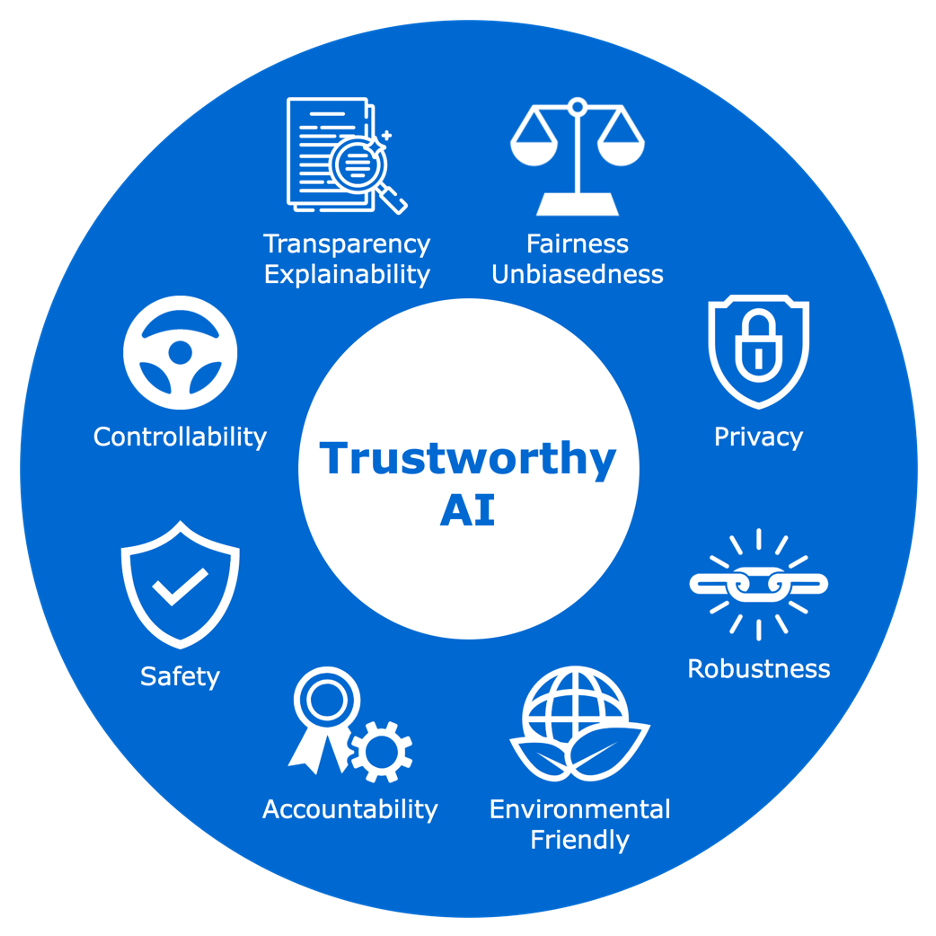}
  \caption{Trustworthiness is a multi-perspective concept in AI and recommender systems.}
  \label{fig:trustai}
  \vspace{-10pt}
\end{figure}

\textbf{Differences from Existing Surveys in Recommendation.}
There have already been several recent surveys focusing on specific ethical topics in recommendation scenarios, such as explainability \cite{zhang2020explainable,chen2022measuring,4401070}, bias and fairness \cite{li2022fairness,wang2023survey,deldjoo2023fairness,chen2020bias,zehlike2021fairness, pitoura2022fairness, zehlike2022fairness, pitoura2022fairness},
privacy protection \cite{huang2019privacy,xu2016privacy,wang2018toward}, attacks and robustness \cite{mobasher2007toward,gunes2014shilling,si2020shilling,mehta2008survey,anelli2022adversarial,deldjoo2021survey}, user controllability \cite{jannach2019explanations, jannach2017user}, etc.
These surveys successfully highlight the importance of the social responsibility of recommender systems, leading to further developments in this important line of research.
However, these topics have only been presented in their own self-contained ways, while a systematic view of trustworthiness in recommendation and the inherent relationship among the various trustworthiness perspectives is highly needed.
The closest work to ours are \citeauthor{dong2020survey} \cite{dong2020survey} and \citeauthor{mobasher2007toward} \cite{mobasher2007toward}. However, \cite{dong2020survey} only covers user social relationship, robustness, and explainability, while \cite{mobasher2007toward} only discusses the attack models and algorithm robustness in recommendation, and they did not examine the internal relationship among these concepts. In contrast, our work introduces trustworthiness over more comprehensive perspectives, highlights the relationship among the perspectives, and sheds light on open problems and future directions to explore the intersection of the perspectives.

\textbf{The Scope of the Survey.}
This survey comprehensively covers over 400 works, encompassing representative papers on the core aspects of trustworthiness such as explainability, fairness, privacy, robustness and controllability in recommender systems published in top Web, IR, RecSys, Data Mining and HCI related conferences and journals. The selected venues include, but are not limited to, IUI, UMAP, CHI, UMUAI, WWW, SIGIR, KDD, RecSys, CIKM, WSDM, ICDM, ICDE, AAAI, IJCAI, ICML, NeurIPS, UAI, TOIS, TORS, TKDE, and TIST, amongst others. Additionally, notable arXiv papers have also been included in this survey.

\textbf{Intended Audience and Paper Organization.}
The primary readers of the survey are RS researchers, technologists, and practitioners aiming to make recommender systems more trustworthy and responsible with their expertise. On the other hand, since the recommender system is a very representative and pervasive human-centered AI system, thus the target readers of the survey also include general AI researchers, practitioners, theorists as well as public policymakers who are interested in trustworthiness, ethics, and regulations of AI.
To achieve this goal, the survey is organized as follows: Section \ref{sec: relationship} provides an overview of different trustworthy recommender system perspectives as well as an introduction to their inherently relationships. Section \ref{sec: rs} introduces the preliminary knowledge of recommender systems and some representative recommendation algorithms. The subsequent sections, namely \ref{sec: explainability}, \ref{sec: fairness}, \ref{sec: privacy}, \ref{sec: robustness}, and \ref{sec: controllability}, delve into explainability, fairness, privacy, robustness, and controllability, respectively. Broadly speaking, this work mainly focuses on addressing trustworthiness issues from the end user's perspective.


\input{sections/relationship}

\input{sections/rs}

\input{sections/explainability}

\input{sections/fairness}

\input{sections/privacy}

\input{sections/robust}

\input{sections/controllability}

\input{sections/conclusion}

\bibliographystyle{ACM-Reference-Format}
\bibliography{main}

\end{document}

%% file: sections/relationship.tex

\begin{table}[t]
\centering
  \caption{Relationship between different trustworthy perspectives. The section number at diagonal positions indicate the corresponding section of each trustworthy recommendation perspective in this survey. The references at non-diagonal positions represent recent studies on cross-perspective relations.}
  \label{tab: relationship}
  \small
  \begin{tabular}[c]{c|p{0.135\textwidth}<{\centering}|p{0.143\textwidth}<{\centering}|p{0.135\textwidth}<{\centering}|p{0.135\textwidth}<{\centering}|p{0.135\textwidth}<{\centering}|}
  \multicolumn{1}{c}{} & \multicolumn{1}{c}{Explainability} & \multicolumn{1}{c}{Fairness} & \multicolumn{1}{c}{Privacy} & \multicolumn{1}{c}{Robustness} & \multicolumn{1}{c}{Controllability}\\ [7pt]
  \cline{2-6}
    \multirow{3}{*}{Explainability} &  & Fairness-aware & Privacy-aware & Robust & Controllable \\
    & Section \ref{sec: explainability} & Explainability & Explainability & Explainability & Explainability \\
    & & \cite{fu2020fairness, zhao2022fairness} & \cite{georgara2022privacy,brunotte2021can} & \cite{slack2021counterfactual,aivodji2020model} & \cite{li2020towards,li2020generate,li2021personalized} \\
    \cline{2-6}
    \multirow{3}{*}{Fairness} & Explainable &  & Privacy-aware & Robust & Controllable \\
    & Fairness & Section \ref{sec: fairness} & Fairness & Fairness & Fairness \\
    & \cite{ge2022explainable,ghosh2022faircanary,zhou2022towards} &  & \cite{liu2022fairness,maeng2022towards} & \cite{solans2021poisoning,zhang2022pipattack} & \cite{wu2022selective,li2021towards} \\
    \cline{2-6}
    \multirow{3}{*}{Privacy} & Explainable & Fairness-aware & & Robust & Controllable \\
    & Privacy & Privacy & Section \ref{sec: privacy} & Privacy & Privacy \\
    & \cite{mosca2021elvira,sanchez2020recommendation} & \cite{ekstrand2018privacy,zhu2022cali3f} &  & \cite{zhang2022comprehensive,wadhwa2020data,bilge2013robust} & \cite{anelli2022user,kelley2008user} \\
    \cline{2-6}
    \multirow{3}{*}{Robustness} & Explainable & Fairness-aware & Privacy-aware & & Controllable\\
    & Robustness & Robustness & Robustness & Section \ref{sec: robustness} & Robustness \\
    & \cite{pawelczyk2022exploring,demontis2019adversarial} & \cite{ovaisi2022rgrecsys,10.5555/3524938.3525692, xu2021robust} & \cite{lam2006you,zhang2021membership,zhang2022comprehensive} &  & \cite{pereira2023the,zhang2019understanding} \\
    \cline{2-6}
    \multirow{3}{*}{Controllability} & Explainable & Fairness-aware & Privacy-aware & Robust & \\
    & Controllability & Controllability & Controllability & Controllability & Section \ref{sec: controllability} \\
    & \cite{tsai2021effects,ngo2020exploring} & \cite{wang2022user} & \cite{wang2019privacy,walter2018user} & \cite{noh2015auro,jambor2012using} & \\
    \cline{2-6}
\end{tabular}
\end{table}

\section{\mbox{Trustworthy Perspectives \& Their Relationships in Recommendation}}
\label{sec: relationship}
{In the extensive literature surrounding trustworthiness, a consensus emerges on its pivotal dimensions: \textit{explainability, fairness, privacy, robustness, and controllability}. These dimensions are central to the effective design and function of Trustworthy Recommender Systems (TRS). It is important to note that these are not the only perspectives of a trustworthy recommender system. However, they are several widely discussed topics in the literature that are deeply connected with trustworthiness of an intelligent system such as recommender system, more specifically:

\begin{itemize}
\item Explainability: When users can not discern the logic behind recommendations, skepticism or doubt may arise, thus hindering users' trust on the recommender system. Such opacity may even escalate to frustration, especially when suggestions appear off-mark. This enigmatic ``black box'' approach may discourage users, making them suspect inherent arbitrariness or bias.

\item Fairness: If users feel that they are not being fairly treated by the system, they eventually lose trust in the system. Without ensuring fairness, the system risks magnifying societal biases. This not only fosters discrimination but also entrenches prevailing stereotypes. It could result in some users or items gaining undue advantage, while others face disadvantages. The platform might face reputation setbacks, prompting boycotts or even legal repercussions.

\item Privacy: If users start to worry about whether their private information is properly protected by the system, they may lose trust in the system. An inadequate emphasis on privacy can pave the way for misuse of user data, exposing personal details. Such oversight may instill a sense of perpetual surveillance among users, dampening their interaction or causing them to leave the platform entirely. Moreover, flouting data protection norms, like the GDPR regulations, invites legal penalties.

\item Robustness: A system lacking robustness is an open target for adversarial attacks, and thus users may lose trust in the system in worrying about being victims of such attacks. Lack of robustness also enables malefactors to skew recommendations -- the system's suggestions may waver in accuracy due to erratic input, eroding user confidence in its reliability.

\item Controllability: If users feel that they are being controlled or even manipulated by the recommender system rather than that they control the system, they ultimately lose trust in the system. Furthermore, in absence of controllability, users may find themselves ensnared in a ``filter bubble'', limited to a homogenized set of suggestions rooted in past interactions. Such a system, resistant to user feedback, may cause dissatisfaction. Users could perceive it as inflexible and unresponsive to evolving tastes.
\end{itemize}

}

To better illustrate the concept of trustworthy recommender system, we explain each perspective of trustworthiness and their relationships with each other. 
We provide a high-level picture of the perspectives and their relationships in Table \ref{tab: relationship}.
We will further expand the discussion of these topics in Sections \ref{sec: explainability}, \ref{sec: fairness}, \ref{sec: privacy}, \ref{sec: robustness}, and \ref{sec: controllability}.

\subsection{Explainability with other Perspectives}
\textbf{Explainability} refers to the ability of RS to provide explanations on the recommendation results to help better understand the inference and reasoning process behind the recommendation models \cite{herlocker2000explaining,4401070,zhang2020explainable}. The recipients of the explanations could be the users who want to better understand the recommended items, the system designers who want to better debug or improve the model, or the policymakers who want to conduct informed decision making \cite{4401070,zhang2020explainable}.
For example, in an e-commerce system, an explanation provided by the RS, such as the item-feature-based explanation, can help to explain why a particular item is being recommended to a specific user. 
Nonetheless, the importance of explainability in RS extends beyond enhancing user understanding---it is closely related to other perspectives of RS trustworthiness. We briefly introduce the relationship between explainability and other trustworthy recommendation perspectives in the following, and we will expand the discussions in Section \ref{sec: explainability}.

\begin{enumerate}
    \item \textbf{Fairness-aware Explainability} aims to generate explanations that are fair. Research works find that the recommendation explanations may be vulnerable to bias and unfairness \cite{fu2020fairness, zhao2022fairness}, e.g., the explanation quality for different user groups could be very different. Therefore, generating fairness-aware recommendation explanations is an important consideration.

    \item \textbf{Privacy-aware Explainability} aims to protect user privacy when generating explanations. Take social recommendation as an example, a frequently used explanation is neighbour-user explanation, which tells the target user that the reason a certain product or advertisement is recommended is because his or her friend(s) clicked this product or advertisement. However, extreme care should be taken when generating such explanations to avoid leaking the private information of these friend(s) because they may not want their online activity to be exposed to others. As a result, how to anonymize such information and meanwhile generating informative explanation is an important problem \cite{georgara2022privacy,brunotte2021can}.
    
    \item \textbf{Robust Explainability} aims to increase the recommender system's robustness to attacks while providing explanations. The reason is because providing explanations may reveal the recommender system's internal working mechanism to outside parties. As a result, if the explanations are leveraged by adversary entities, it may expose the system to attacks by helping the adversary entities to create more targeted attacking methods. One example is counterfactual explanation \cite{ghazimatin2020prince,tran2021counterfactual,kaffes2021model,tan2021counterfactual} which is especially vulnerable to malicious usage, because it directly tells users what kind of interventions can change the recommendation result in what ways \cite{slack2021counterfactual,aivodji2020model}. Therefore, it is crucial to develop robust explainable recommendation systems.

    \item \textbf{Controllable Explainability} aims to improve user controllability in the explanation generation process from different perspectives, such as the level of detail and the level of expertise. 
    One example is user-controllable natural language explanation generation \cite{li2020towards}, which provides users with the flexibility of choosing the target feature(s) to explain over an item \cite{li2020generate,xie2023factual,li2021personalized}. This is because practical recommender systems and explanations usually involve a large amount of features, making it difficult and time-consuming for users to go through each and every feature of an item. As a result, controllable explanation allows the system to reveal parts of the explanation that are relevant to the users' interest, providing different levels of detail for users. Users may also control the expertise level of the explanations, i.e., if users are experts on recommender systems, then they may want complex and complete explanations, otherwise, they may want simple or partial explanations, allowing level-of-expertise controllability.
    They might also desire greater control over the length and the sentiment of the explanation, which requires joint consideration of controllable generation \cite{zhang2022survey} and explainable recommendation \cite{zhang2020explainable}.
    It should be emphasized that controllable explanation does not mean that users can manipulate the system to generate fake explanations.
    Instead, to assist users' decision-making process, such explanations should provide faithful facts about the item, which can be retrieved from the item database \cite{li2021extra} or generated by language models followed by a fact-checking process \cite{xie2023factual}. Moreover, it is also important to use unambiguous and user-friendly controllable explanation interfaces in real-world systems to ensure that the system does not mislead users. More detailed discussions are provided in Section \ref{sec:relationship_explainability}.
    


\end{enumerate}

\subsection{Fairness with other Perspectives}
\textbf{Fairness} aims to mitigate the bias and unfairness in recommender systems \cite{deldjoo2023fairness,wang2023survey,li2022fairness}. 
For instance, in online job marketplaces, recommender systems may lead to racial or gender discrimination by disproportionately recommending low-paying jobs to certain user groups due to echoing the existing bias in the recorded training data \cite{Geyik2019}. The research of fairness in recommender systems considers multiple stakeholders, such as the users, the platform, and the policy makers \cite{abdollahpouri2019multi}. Besides, there also exist various types and different definitions of fairness. In Section \ref{sec: fairness}, we will introduce the fairness of recommender systems in greater detail. In this subsection, we briefly discuss the relationship between the fairness perspective and other trustworthy perspectives so as to help readers build a broad overview.

\begin{enumerate}
    \item \textbf{Explainable Fairness} aims to generate explanations so as to help users, system designers or policy makers to understand why the recommender system is fair or unfair \cite{ge2022explainable,ghosh2022faircanary,zhou2022towards}. Explainable fairness is especially important for recommender systems because real-world recommender systems usually make decisions based on huge amounts of both explicit and implicit features. As a result, the features that lead to unfairness problems may not be those obvious features such as gender, race or age in other intelligent decision making systems such as college admission \cite{kleinberg2018algorithmic} and parole decision \cite{corbett2017algorithmic}. Because of this, it is usually very difficult for users, system designers or policy makers to manually decide which feature(s) lead to unfairness problems \cite{ge2022explainable}. This makes explainable fairness an important task which automatically probes and explains the unfairness problem of a system.

    { Moreover, introducing fairness into the recommendation objective could lead to recommendations that seem unusual or unexpected from a strictly personalization-driven viewpoint. As such, it becomes imperative to communicate the presence and purpose of this fairness objective to the end users. When users understand that the system is taking steps not just to cater to their personal preferences but also to uphold values of fairness, it can enhance users' trust in the platform.}
    \item \textbf{Privacy-aware Fairness} aims to guarantee system fairness in a privacy-preserving manner. This poses unique challenges in recommender systems for two reasons: 1) the system may need access to users' sensitive features so as to increase the fair treatment for disadvantaged users; however, 2) users may be unwilling to share their sensitive features with the system for privacy concerns. Sometimes, the reason that users are unwilling to share their sensitive features is exactly because they worry about being unfairly treated by the system, even though the system aims to increase their fairness, which causes a dilemma for recommender systems. One approach to solving the privacy-aware fairness problem is federated fairness \cite{liu2022fairness,maeng2022towards,zhu2022cali3f}, which adopts the federated learning framework for fairness so that users can keep their sensitive information on their personal device and do not have to explicitly share such information with the model.

    \item \textbf{Robust Fairness} aims to develop fairness-aware recommendation algorithms that can withstand intentional or malicious attacks in the recommendation process \cite{solans2021poisoning,zhang2022pipattack,deldjoo2021survey,anelli2021adversarial}. Similar with attacks on the recommendation performance where certain entities intentionally promote the ranking of their own items, recommendation fairness may also be vulnerable to attacks where certain individuals or groups of users may disproportionately promote their recommendation quality or the exposure opportunity of their items. A good recommender system should not only be fair but also be resistant to such attacks on its fairness.

    \item \textbf{Controllable Fairness} is to offer users an autonomy over the features they consider essential for fair treatment. It is not about turning the system's fairness mechanism on or off, but rather allowing users to prioritize which aspects of fairness they value most for their own recommendation list, based on their unique life experiences and perspectives. For example, some users may be sensitive to gender and do not want their recommendations to be discriminated by this feature, while others may care more about the age feature and are less concerned about gender. As a result, it is important to give users the ability to control which feature(s) they are particularly sensitive about, and thus the system can adjust their own recommendation list accordingly \cite{wu2022selective,li2021towards}. Meanwhile, controllable fairness implementation needs to take extreme care so that users' personal biases do not abruptly interfere with the fairness mechanism of the whole system, and a proper fairness-utility relationship is achieved when users adjust their fairness demands, which we will discuss in greater details in the following section corresponding to fairness.
\end{enumerate}


\subsection{Privacy with other Perspectives}
\textbf{Privacy} in the context of recommender systems aims to protect users' private data from being disclosed, leaked or attacked. The goal is to ensure that users' sensitive information, such as personal preferences, interaction history, or demographic information, is kept confidential and only used for providing personalized services to the user. Furthermore, users also have the right to tell the system that they do not want certain information to be used in their personalized service, or choose to opt out of personalized services at all, such as the right to the forgotten on the web \cite{rosen2011right}, and the system should be able to accommodate users' requests in terms of their private data accordingly. Section \ref{sec: privacy} will discuss privacy of recommender systems in detail. In this subsection, we elaborate on the relationship between privacy and other trustworthy recommendation perspectives.

\begin{enumerate}
    \item \textbf{Explainable Privacy} aims to explain how the recommender system guarantees privacy, why it is effective against certain privacy risks, what are possible side effects, and why the suggested privacy setting is optimal for users. For example, users sometimes have to trade-off between the amount of shared information and the quality of personalized service---when the user prefers more privacy and shares less information with the recommender system, the system's personalization ability may decrease; on the other hand, if the user shares more information, although the personalization quality may increase, the user may have to bear higher risks of privacy concerns. As a solution, the system can recommend the optimal privacy setting tailored to each user so that users can achieve the best personalized service while sharing the least amount of information \cite{yu2018leveraging,sanchez2020recommendation}. Meanwhile, the system can explain to users why the recommended setting is optimal \cite{mosca2021elvira}. Furthermore, if using certain information is inevitable to provide personalized services for users, the system can explain to users why it needs access to the information so as to gain users' trust \cite{kunkel2019let}.

    \item \textbf{Fairness-aware Privacy} aims to promote privacy protection in a fair manner. This can be interpreted on two dimensions: 1) The system's privacy scheme should provide fair and nondiscriminatory privacy protection to all users irrespective of their group membership or level of vulnerability, i.e., the privacy regimes should not disproportionately fail to protect certain users which may result in disparate impact on the effectiveness of privacy protection \cite{ekstrand2018privacy}; 2) The system should not lead to unfair service quality due to its privacy protection regime.
    For example, some users are concerned about their gender information and thus do not share gender-related information with the system; meanwhile, some other users may be more concerned about their age information and thus reserve their age-related information from the system. This may lead to privacy-induced bias in the training data, i.e., on each feature, the data will be dominated by those users who are willing to share this feature \cite{fernando2021missing,zhang2021assessing,noiret2023fairly}. 
    This further leads to fairness issues in recommendation since the recommendation performance for one user group would be biased by another user group who choose to share the related information with the system \cite{zhu2022cali3f,ekstrand2018privacy}. As a result, fairness should be taken into account when developing privacy protection policies \cite{resheff2018privacy,ekstrand2018privacy}.


    \item \textbf{Robust Privacy} aims to guarantee that the recommender system's privacy protection mechanism is resistant to attacks, such as data poisoning attacks \cite{wadhwa2020data}, attribute inference attacks \cite{zhang2022comprehensive} and profile injection attacks \cite{bilge2013robust}. In certain systems such as decentralized recommendation, there may even exist inevitable trade-off between robustness and privacy and thus we have to consider the best balance between the two \cite{cheng2009trading}. For example, federated architecture helps to protect user privacy by keeping user's private data on the user's own device. However, due to various reasons such as users' weak passwords, mis-configuration or poor usage habits, on-device data could be easier to attack for malicious parties compared to on-cloud data which is protected by professional data scrutiny team. As a result, careful consideration should be taken for the robustness of privacy protection mechanisms.

    \item \textbf{Controllable Privacy} provides users with the ability to control the type and amount of personal information that can be used by the system in generating recommendations \cite{anelli2022user,kelley2008user}. For example, users can control the type and amount of sensitive data that leaves their devices in a federated recommender system \cite{anelli2022user}. This allows users to explore the trade-off between privacy and service qualify and find out the sweet spot by themselves. 
\end{enumerate}



\subsection{Robustness with other Perspectives}
\textbf{Robustness} refers to the ability of a recommender system to maintain its performance and effectiveness even in the presence of various forms of challenges or adversarial attacks. These challenges can take many forms, such as data poisoning, model evasion, or adversarial examples. We will expand the robustness of recommender systems in Section \ref{sec: robustness}. In this subsection we discuss its relationships with other trustworthy perspectives in recommender system.

\begin{enumerate}
    \item \textbf{Explainable Robustness} aims to provide explananations to users or system designers about why the system is resistent to attacks, or why certain component of the system is vulnerable to attacks and thus needs careful attention or improvement \cite{demontis2019adversarial}. The recommendation model's explainability may also help system designers to develop better defense strategies. For example, counterfactual explanation can help system designers to understand how the system would respond to possible data poisoning or profile injection attacks \cite{pawelczyk2022exploring} so that system designers can develop preventive strategies in advance.
    \item \textbf{Fairness-aware Robustness} aims to guarantee that different users or user groups are fairly protected from attacks or fairly influenced in face of possible attacks \cite{ovaisi2022rgrecsys}. For example, users with fewer historical interaction records may experience more significant drop in recommendation quality when the system is under profile injection attack, since their recommendation can be easier to influence by injecting profiles \cite{gunes2014shilling}. As a result, it is important to explore how users can be fairly protected under attacks. Furthermore, existing works on machine learning (ML) have shown that adversarial robustness and fairness could negatively affect each other during training \cite{10.5555/3524938.3525692, xu2021robust}. It is possible that similar trade-offs between adversarial robustness and fairness may also exist in recommender systems, which is an area remaining largely unexplored.
    


    \item\textbf{Privacy-aware Robustness} aims to protect user privacy when possible attacks happen to the recommender system. Though most attacks to recommender systems aim to manipulate the recommendation results and promote target items in the recommendation list, some attacks simply aim at stealing users' private information from the system such as membership inference attacks \cite{zhang2021membership,wang2022debiasing,yuan2023interaction} or attribute inference attacks \cite{zhang2022comprehensive,zhang2021graph,zhang2020practical,chai2022efficient,aktukmak2019quick}. Besides, certain attacks may probe users' private data out of the system as a byproduct. For example, when the recommender system provides neighbour-user-based explanations for the injected profiles, the attacker may be able to infer the real users' information from the system \cite{coba2022recoxplainer,sharma2013social}. As a result, robustness research should also make sure users' information is protected from possible attacks \cite{lam2006you}.
    
    \item \textbf{Controllable Robustness} is essential in system design, and should not be user-driven to prevent adversarial behaviors. The concept of ``control'' in recommender systems primarily pertains to system designers, allowing system designers to generate controllable attacking examples or malicious activities to detect the weaknesses of the system and to improve the robustness of the system \cite{pereira2023the,zhang2019understanding}.
    
\end{enumerate}

\subsection{Controllability with other Perspectives}
\textbf{Controllability} 
enables users to have better control and customization over the recommendations they receive. This is accomplished by providing users with a range of feedback mechanisms, allowing them to set preferences and interact with the recommendation results \cite{jannach2019explanations, jannach2017user}. By offering greater controllability, these systems can enhance user satisfaction and trust, as well as improve the accuracy and relevance of recommendations. We will discuss controllability in Section \ref{sec: controllability} and we briefly introduce its relationship with other trustworthy recommendation perspectives in the following.

\begin{enumerate}
    \item \textbf{Explainable Controllability} aims to explain to users how and why their preferences or interactions will help them to steer the system towards certain directions \cite{tsai2021effects,ngo2020exploring}.
    This is important because controlling an intelligent system without even knowing the underlying mechanism could be harmful or even dangerous.
    Explainable controllability helps to explain to users how their actions will lead to certain outcomes so that they know the potential consequences before really implementing the control.
    
    \item \textbf{Fairness-aware Controllability} aims to mitigate bias and unfairness of the recommender system through user controllability and to prevent the system from introducing new bias or unfairness issues when users control the system. For example, users may be trapped in filter bubbles when interacting with the recommender system, which may further lead to unfairness issues due to users' frequent clicks of popular items. By providing users with the option to control the system's recommendations, it can help users to escape from the bubble, which further contribute to the fairness of the recommender system \cite{wang2022user}. 
    However, it is also important to note that users may not share the same fairness goals with the system designers and that users' control of a system may prevent the system from achieving fairness. As a result, system designers may need to take users' influence into account when designing fairness policies.
    

    \item \textbf{Privacy-aware Controllability} aims to protect users' or item providers' privacy when users control the recommender system. While providing controllability can improve users' experience, controllability may also introduce additional privacy risks. For example, enhancing the controllability of a recommender system may require the collection of additional information from the users. Additionally, it becomes easier for adversary users to probe the private information from the system, especially when the ``items'' being recommended are also human users such as in friend recommendation \cite{shu2017user} and advisor recommendation \cite{rahdari2021connecting}, because users have better ability to manipulate and compare the recommendation results by specifying various controllability options. For example, users may control the system to recommend male or female friends for them, and the system may return such results accordingly, which may unwillingly expose the private gender information of the recommended users.
    Therefore, it is crucial to explore controllable recommender systems with privacy protection \cite{wang2019privacy,walter2018user}.


    \item \textbf{Robust Controllability} aims to guarantee that the controllability function of the recommender system is robust to possible attacks. Recommender systems with controllability face unique challenges in terms of possible attacks or disturbances. For example, adversarial parties may manipulate the system based on data injection so as to lure the users to click manipulated items when users control the system \cite{jambor2012using}. As a result, it is important to make sure that the controllability functions are robust so that users can safely control the system towards their expected directions \cite{noh2015auro}.


\end{enumerate}

%% file: sections/rs.tex
\section{Recommender System Basics}\label{sec: rs}
{Recommendation accuracy remains central to the efficacy of trustworthy recommender systems. We now subsequently outline the fundamental research problem associated with personalized recommendation and present a formalized representation of recommender systems.}


\subsection{Input and Output of RS}

\begin{figure}[t]
  \centering
  \includegraphics[width=0.8\textwidth]{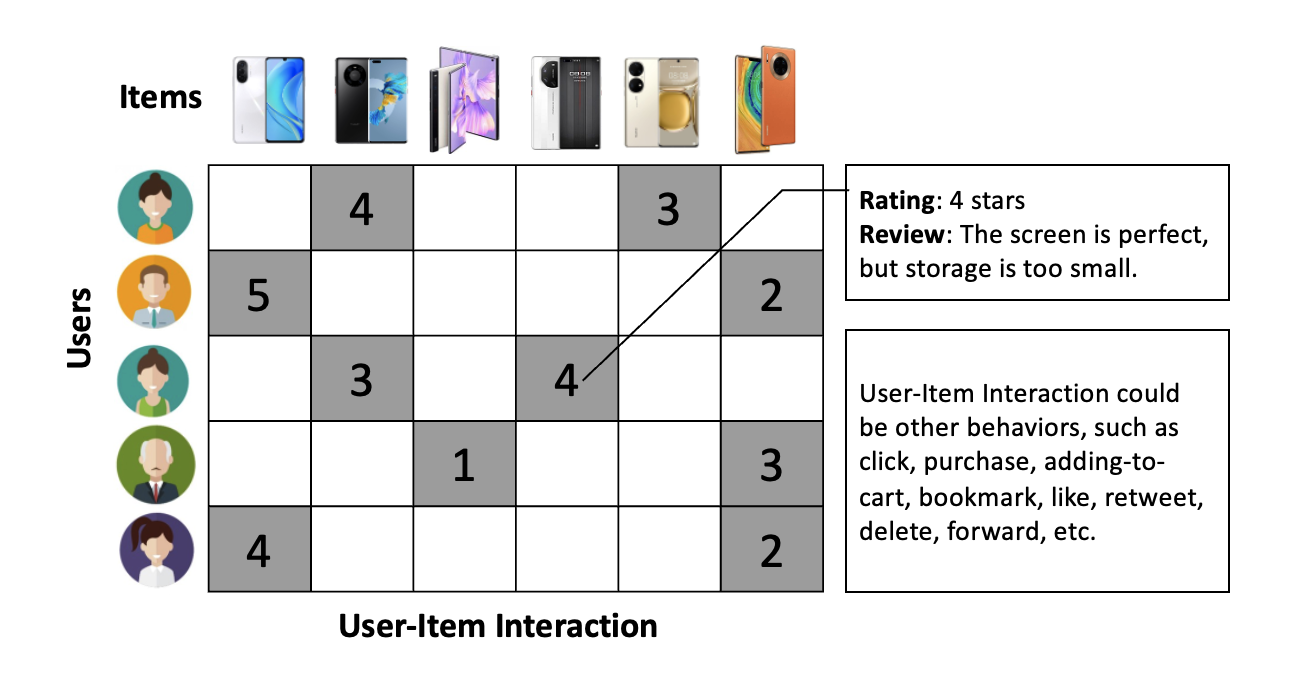}
  \vspace{-10pt}
  \caption{Typical inputs of a recommender system.}
  \label{fig:input}
  \vspace{-10pt}
\end{figure}

Recommender systems may have various forms of possible input data. A basic recommender system usually has three types of input data: user, item, and interaction, where the interaction could be users' click, purchase, rating, review, or other behaviors over the items (Figure \ref{fig:input}). 
We should notice that the ``item'' here has a very broad meaning and could be various types of things. For example, it could be products on an e-commerce shopping website, tweets or ads on social networks, hotels or air tickets on trip planning websites, videos or music in media streaming applications, jobs in online marketplaces, or even other users of this website such as in friend recommendations. 
Different types of items have different attributes. For example, products of shopping sites may have brands, manufacturers, sizes, and weight, while for media recommendation, the attributes may include the type of media, style, and content descriptions, and trip recommendation may focus on travel methods, time duration, stop numbers and estimated cost, etc. Besides, the ``user'' may not only be an ID in the recommender system, but also a profile that can describe the user. 
User profiles can have distinct forms in different application scenarios or in different recommendation algorithms. An intuitive and understandable form of the user profile is the registration information of the user, such as age, gender, annual income, location, and active time intervals.

Apart from users and items, interaction is an important input of recommender system as well, which is the connection between users and items. One of the most commonly used types of interaction is the rating information from users to items.
The 5-star rating scale is widely used on major websites, which reflects the user's preference for the item. Rating is denoted as an integer between 1 and 5 (1 and 5 included) in most recommendation algorithms. The preference of users may also be revealed by other interactions, such as users' reviews, clicks, and purchase histories. Usually, we can separate these contents into two types, explicit feedback, and implicit feedback. Explicit feedback such as ratings and reviews is collected when users actively and explicitly tell the system about their preferences for an item, while implicit feedback such as user clicks is passively recorded when users interact with the website interface, and the data can implicitly reflect the users' preference.
For instance, if a user explores a lot of detailed information about an item or the user spends a lot of time viewing the web page of the item, then it implies that the user may have a strong interest in the item.

The output of a recommender system usually includes a personalized recommendation list tailored to the user and the explanations accompanying the recommendations (Figure \ref{fig:output}). The process of generating the output usually includes three stages: Predict, Rank, and Explain (PRE).
The predict stage aims to make preliminary predictions about the user's preferences on the items and generate a preliminary list of candidate items for the user. The preliminary list could be very long such as thousands or more, as a result, this stage is sometimes also called the ``recall stage'' in industrial production environments. The ranking stage refines the list from the previous predict stage and generates the final recommendation list for the user, which usually includes just a few or a few tens of items. Although the easiest way of ranking is just sorting the items according to the predicted scores in the predicting stage, a lot of other important factors play a role in this stage, such as refining the preference predictions based on more information and more meticulously designed algorithms that can only be executed on the smaller amount of items selected by the previous stage, re-ranking the list based on diversity, fairness, serendipity, business or other considerations, and responding to user's real-time requests. Finally, the explain step generates explanations for each recommended item to justify the recommendation. Depending on the explainable recommendation method used to generate the explanations, the explain step may either happen simultaneously with the ranking step or after the ranking step: if the ranking model used in the ranking step is intrinsically explainable, then the model can directly generate explanations alongside the recommendation list, and thus the explain step happens at the same time; otherwise, if the ranking model is black-box, then a post-hoc explanation generation model is needed in the explain step to generate the explanations after the recommendation list is provided.

\subsection{Representative Recommendation Algorithms}

There are a diverse set of recommendation algorithms. In this survey, we broadly categorize the recommendation algorithms into three stages: shallow models, deep models, and big models.

\begin{itemize}
\item \textbf{Shallow Models}: The pioneer works of recommender systems started with shallow models which take expert-designed similarity functions to extract simple and effective user-item matching patterns from data. This includes both collaborative filtering approaches \cite{resnick1994grouplens,konstan1997grouplens,sarwar2001item,Linden2003} and content-based filtering approaches \cite{balabanovic1997fab,van2000using}. 
Collaborative Filtering (CF) is one of the most fundamental approaches for recommender systems, which has been widely used in real-world systems.
Early CF methods adopted simple similarity functions such as cosine similarity and inner product for matching.
The basic idea is that similar users may share similar interests and similar items may be liked by similar users. CF methods can be further divided into two groups according to whether user-item similarity is calculated based on direct user/item features or learned user/item features. Direct-feature-based CF is also called memory-based CF \cite{burke2002hybrid}, which calculates user similarity or item similarity for recommendation based on a similarity measurement over users' and/or items' direct historical data. For example, User-based CF \cite{resnick1994grouplens, konstan1997grouplens} directly takes each row vector in the user-item rating matrix as the representation for each user, and calculates the user similarity based on Pearson correlation coefficient or cosine similarity;
Item-based CF \cite{Linden2003,sarwar2001item} takes the column vector of the user-item rating matrix as the representation for each item, and also employs similarity functions to calculate the similarity between items for recommendation. 

Learned-feature-based CF is also called model-based CF \cite{burke2002hybrid}, which adopts a model to learn the user and item representations for making predictions. Such models are also frequently named Latent Factor Models (LFM). For example, Matrix Factorization (MF) \cite{svd} learns each user and item as a latent embedding vector and takes the inner product between the user and item vectors to calculate the user-item similarity for recommendation; Probabilistic Matrix Factorization (PMF) also learns user and item latent representations for recommendation and provides a probabilistic interpretation for matrix factorization under the Bayesian learning framework \cite{mnih2008probabilistic}. 
Content-based filtering approach, on the other hand, relies on the various user and item content features such as user profiles and item descriptions for recommendation \cite{balabanovic1997fab,van2000using,pazzani2007content, bobadilla2013recommender}. 
For example, a movie can be described by its features such as genre, director, cast, length, and language. Content-based filtering calculates the similarity between items based on their content similarity (e.g., same director, overlapping cast) and recommends those movies to a user that are contently similar to the user's previously watched movies \cite{balabanovic1997fab,van2000using,pazzani2007content,kompan2010content}.

\begin{figure}[t]
  \centering
  \includegraphics[width=1\textwidth]{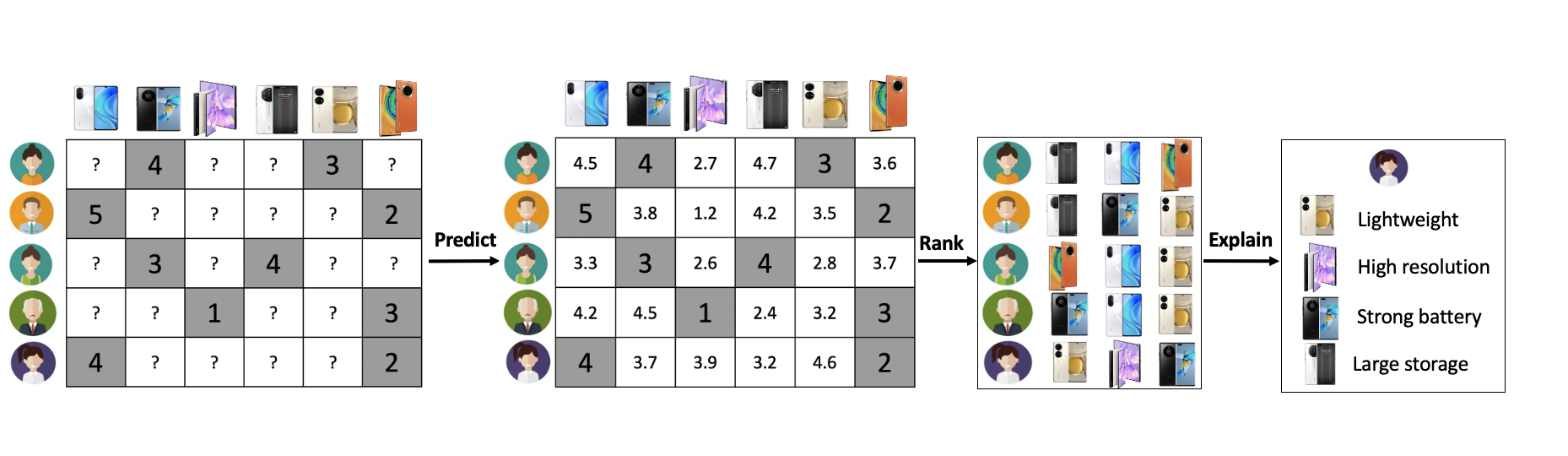}
  \vspace{-25pt}
  \caption{Typical workflow and output of a recommender system.}
  \label{fig:output}
  \vspace{-10pt}
\end{figure}

\item \textbf{Deep Models}: The development of deep learning and neural networks have further improved recommendation methods \cite{zhang2019deep}. The various neural approaches to recommendation can be broadly classified into two categories: Collaborative Filtering (CF) \cite{cheng2016wide,xue2017deep,zhang2017joint,zheng2017joint,xu2021causal,xu2022dynamic,xu2021deconfounded} and Collaborative Reasoning (CR) \cite{chen2021neural,shi2020neural,chen2022graph,chen2022learn,zhang2022neuro,zhang2022graph,yuan2021probabilistic}, while the CF approach can be further classified into the similarity learning methods \cite{cheng2016wide,he2017neural,hsieh2017collaborative,xue2017deep} and the representation learning methods \cite{zhang2017joint,ai2018learning,mcauley2015image,zheng2017joint,xu2021causal}.
Deep learning based CF considers recommendation as a perceptive learning problem, which employs similarity learning or representation learning to extract perceptive correlation patterns from data for matching and recommendation; while deep learning based CR considers recommendation as a reasoning problem, which employs logical reasoning or causal reasoning for user behavior prediction and recommendation. More specifically, for deep learning based CF, the similarity learning approach adopts simple user/item representations (such as one-hot vector) and learns a complex matching function (such as a neural prediction network) to calculate user-item matching scores \cite{cheng2016wide,he2017neural,hsieh2017collaborative,xue2017deep}, while the representation learning approach learns rich user/item representations from text, image, knowledge, etc. and adopts a simple matching function (e.g., inner product) for efficient matching score calculation \cite{ai2018learning,mcauley2015image,zhang2017joint,zheng2017joint}. It is worth noting that there exist debates over whether complex matching functions are better than simple functions \cite{dacrema2019we,dacrema2021troubling,ferrari2020critically,rendle2020neural}. 
The above recommendation algorithms use a user's historical interaction to learn the static preference, however, in many real-world scenarios, the next behavior of a user not only depends on the static long-term preference but also relies on the current intent \cite{fang2020deep}. Therefore, sequential recommendation (also related to session-based or session-aware recommendation), which models the user behavior as a sequence for future behavior prediction, has become increasingly important in academia and industry \cite{hidasi2015session,chen2018sequential,tang2018personalized,sun2019bert4rec,kang2018self,liu2018stamp,huang2018improving,ma2019hierarchical,zhou2020s3,de2021transformers4rec}. Traditional sequential recommendation models employ simple machine learning approaches to model sequential data, such as Markov Chain \cite{rendle2010factorizing} and Session-based KNN \cite{hu2020modeling}. With the development of deep learning techniques, many deep models obtain tremendous achievements in the sequential recommendation, including RNN \cite{hidasi2015session}, CNN \cite{tang2018personalized, yuan2019simple}, LSTM \cite{wu2017recurrent}, BERT \cite{sun2019bert4rec,de2021transformers4rec}, attention models \cite{wang2018attention,kang2018self}, and memory networks \cite{chen2018sequential}. Another important direction is learning to rank for recommendation \cite{bpr,he2016vbpr,abdollahpouri2017controlling}, which learns the relative ordering of items instead of the absolute preference scores. A representative method for learning to rank in recommendation is Bayesian personalized ranking (BPR) \cite{bpr}, which is a pair-wise learning-to-rank method. It is also further generalized to take other information sources such as images \cite{he2016vbpr}.

\item \textbf{Large Models}: Recently, Large Models such as Large Language Models (LLM) have achieved surprisingly good performance in many AI sub-fields, which have the advantages of emergent capabilities from model size, extracting useful information based on self-supervision, unifying various downstream tasks based on pre-training, fine-tuning and prompting, as well as generalizing to zero-shot or few-shot cases \cite{bommasani2021opportunities,wei2022emergent}. 
Many powerful language models have been developed for natural language tasks such as T5 \cite{raffel2020exploring}, GPT-3 \cite{brown2020language}, OPT \cite{zhang2022opt}, PaLM \cite{chowdhery2022palm}, BLOOM \cite{scao2022bloom} and LLAMA \cite{touvron2023llama}, which show impressive performance on language understanding, generation and reasoning tasks and also supported successful applications such as ChatGPT. Since language grounding is a powerful medium that can describe almost any data, problem, or task, many other problems can be formulated as language sequences by learning the corresponding token embeddings and integrating them with normal word embeddings. For example, many pre-trained vision-language models based on visual token embedding have demonstrated strong power in visual understanding, generation, and reasoning tasks as well as vision-language co-learning tasks, such as CLIP \cite{radford2021learning,geng2023hiclip}, DALL-E \cite{ramesh2021zero,ramesh2022hierarchical} and GPT-4 \cite{openai2023gpt4}. 

Recommender system research has demonstrated a similar trend, besides, since personalization is one of the most unique and important characteristics of recommender system research, the recommender system community has been leading the research on Personalized Foundation Models (PFM) by reformulating various recommendation tasks into unified language tasks \cite{geng2022recommendation,li2023large}. For example, \citet{geng2022recommendation} introduce P5, a Pre-train, Personalized Prompt, and Predict Paradigm for recommendation, together with its variants such as VIP5 \cite{geng2023vip}, UP5 \cite{hua2023up5}, P5-ID \cite{hua2023index}, OpenP5 \cite{xu2023openp5}, and GenRec \cite{ji2023genrec}, they unify various recommendation tasks under a language understanding and generation framework based on personalized prompts. P5 serves as a foundation model for various downstream recommendation tasks and leads to a universal recommendation engine. \citet{muhamed2021ctr} propose CTR-BERT, which fine-tunes a pre-trained BERT model \cite{devlin2018bert} for click-through-rate (CTR) prediction. Meanwhile, \citet{sileo2022zero} adapt GPT-2 \cite{radford2019language} for similar item recommendation through zero-shot prediction. \citet{li2022personalized,li2021personalized} develop PETER and PEPLER, a Transformer-based architecture that incorporates discrete and continuous prompt learning to generate explainable recommendations. \citet{cui2022m6} present M6-Rec, which utilizes an industrial pre-trained language model to convert multiple recommendation tasks into language understanding or generation. 

Compared to deep recommendation models, large recommendation models possess superior generalization abilities and exhibit zero-shot learning capabilities due to their ability to learn diverse forms of knowledge from various tasks and domains. Furthermore, owing to their robust generalization capabilities, one can fine-tune a pre-trained LLM or retain a frozen model with prompt learning for downstream recommendation tasks. 
{ Yet, current LLM-based recommenders still have several limitations, including their limited ability to offer personalized suggestions, their tendency to recommend fictitious items \cite{bender2021dangers,wang2023survey}, and a significant demand for extensive computational resources \cite{bender2021dangers}.
}
\end{itemize}

In the upcoming sections, we will explore ways to incorporate trustworthiness considerations into recommendation models and trustworthy research directions for recommender systems. Specifically, we will delve into several perspectives, including explainability, fairness, privacy, robustness and controllability as well as the relationships between these perspectives in the development of trustworthy recommender systems.

%% file: sections/explainability.tex
\section{Explainability}\label{sec: explainability}
{
Explainable recommendation has been an important area in both industry and academia, aiming to improve transparency, user satisfaction, and trustworthiness over the recommender systems \cite{4401070,zhang2020explainable,zhang2014explicit,bilgic2005explaining}. Prior studies have demonstrated the potential for appropriate explanations to improve user satisfaction of the recommendations \cite{herlocker2000explaining, tintarev2010designing}, as well as enhance user experience in various other dimensions, such as system transparency, trust, effectiveness, efficiency, and scrutability \cite{bilgic2005explaining, cramer2008effects, tintarev2010designing}.
Specifically, the goal is to provide comprehensible justifications along with recommended items to help stakeholders make better decisions and simultaneously improve the transparency and trustworthiness of recommender systems.
In addition to benefiting stakeholders, explanations in recommender systems can also assist model developers to understand and debug the decision-making process so as to enhance end-user engagement and trustworthiness \cite{4401070}. 
}

\subsection{Overview of Explainable Recommendation}
The topic of explainable recommendations has been a subject of research for over two decades \cite{herlocker2000explaining,bilgic2005explaining,pu2006trust,10.1145/1502650.1502661, zhang2014explicit, zhang2020explainable}. The increasing attention from both academia and industry indicates the growing recognition of the importance of explainable recommendations in various domains \cite{zhang2020explainable,4401070}. 
Existing explainable recommendation researches can be broadly classified in terms of two dimensions corresponding to the human-computer interaction (HCI) perspective and the machine learning perspective of explainable recommendation research \cite{zhang2020explainable}: the information sources to present explanations and the models to generate explanations. The dimension of information sources or display styles of the explanations are more studied from the HCI perspective to understand the effectiveness of various forms of explanations such as textual sentence explanations and visual explanations. The dimension of explainable models focuses on how to generate these explanations from the machine learning perspective of explainable recommendation research. 
As an important sub-field of AI and machine learning research and due to the fact that recommendation naturally involves humans in the loop, the recommender system community has been leading the research on Explainable AI ever since, which triggers a broader scope of explainability research in other AI and machine learning sub-fields \cite{zhang2020explainable,confalonieri2021historical},
such as explainability in scientific research \cite{li2022from,tan2023explainablefold}, computer vision \cite{wu2021deconfounded}, natural language processing \cite{oussalah2021ai,pugoy2020bert,cai2022multi,geng2022improving,li2021personalized}, graph neural networks \cite{tan2022learning,xian2021neural}, database \cite{weikum2021machine,glavic2021trends}, healthcare systems \cite{zucco2018explainable,halder2017health,porat2020public}, online education \cite{umemoto2020toward,al2021first,barria2019making,takami2022educational,ooge2022explaining}, psychological studies \cite{tesic2020explanation} and cyber-physical systems \cite{himeur2021survey,sardianos2021emergence,himeur2021artificial,alfrink2022tensions,andric2021climbing}.

Explainable recommendations provide additional explanations on the predicted results to better understand the inference and reasoning process behind black-box prediction models.
As a crucial part of the modern paradigm AI, deep neural networks have increasingly contributed to the development of most state-of-the-art machine learning systems. 
However, these deep neural models are treated as black-box since they are too complicated and opaque to understand \cite{zhang2019deep,gao2020deep,du2019techniques,xu2019explainable}. 
Most of them are not designed toward explainability and transparency, which leads to negative consequences in many human-centered scenarios \cite{gunning2019darpa,molnar2020interpretable}.
For instance, if a medical diagnosis system fails to provide supportive evidence to justify why the prediction is correct or not, the doctor can hardly adopt the automated decision even if its actual accuracy is high. The same situation can be applied to other domains such as e-commerce \cite{xian2021ex3} and digital marketing \cite{xian2021exacta}. The fundamental goal of explainability in AI is to open up the ``black-box'' of the pipelines in AI-related domain, not only to provide trustworthy explanations to users, but also to drive toward more interpretable models. As an important type of intelligent decision-making system, the modern recommender system is expected to provide high-quality recommendation results as well as personalized and intuitive explanations with better user engagement, which are important for many practical applications such as e-commerce and social media platforms.

\subsection{Different Styles of Explanations in Recommender Systems}

The use of accurate and understandable explainable recommendation algorithms is crucial for enhancing the reliability and trustworthiness of RS. However, it is important to note that the style of explanations can often play a significant role along with the algorithm itself in improving the user experience of explainable recommendations. To this end, previous research has employed various display styles to illustrate the rating distribution and advantages/disadvantages of recommendations, such as statistical histograms or pie charts \citep{herlocker2000explaining, bilgic2005explaining, tintarev2007effective, mcsherry2005explanation}. More recent approaches include visual analytics, as proposed by \citet{du2019eventaction}, which aim to provide users with a better understanding of the underlying reasoning behind recommendations, ultimately leading to improved user satisfaction and engagement. These findings highlight the importance of providing clear and concise explanations to users to enhance the overall effectiveness of RS.

In the context of collaborative filtering based recommendation models that rely on users' feedback, using relevant users or items as explanations can be effective. For example, \citet{herlocker2000explaining} compared the effectiveness of different display styles for explanations in user-based collaborative filtering, such as an aggregated histogram of the neighbors' ratings or the detailed ratings of the neighbors. In item-based collaborative filtering \citep{sarwar2001item}, explanations can be provided by informing the user that the recommended item is similar to other items the user has liked before. Additionally, \citet{tintarev2007explanations} developed a prototype system to study the effect of relevant-user and relevant-item explanations in order to better understand how explanations can improve recommender systems.

In contrast, content-based recommendation models generate recommendations based on the matching between the user profile and the item feature. Therefore, it is straightforward to provide explanations based on features \citep{pazzani2007content}. These features can come from either users or items. For instance, \citet{10.1145/1502650.1502661} used tags as features for explanations and displayed tag relevance by bar charts and tag preference by rating stars. Other researchers have employed visual aids such as radar charts, as in \citet{hou2019explainable}, or demographic features, as in \citet{zhao2014we} and \citet{zhao2016exploring}, to improve the explainability of recommendations.

Another popular display method is providing explanation sentences to users, which can be either template-based or generation-based. For example, \citet{wang2018explainable} used template-based explanations based on feature and opinion words, while \citet{costa2018automatic} utilized long-short term memory (LSTM) to generate explanation sentences based on user reviews. Recently, \citet{li2022personalized,li2021personalized} developed large language models to generate explanations. Finally, for recommendation models utilizing visual images, visual explanations such as region-of-interest highlights \citep{chen2019personalized} or taking pixels of the images as explanations \citep{lin2018explainable} can be particularly intuitive and effective.

\subsection{Explanation Generation in Recommender Systems}
In this section, we first discuss the distinctiveness of explanations generated for recommender systems compared to general explainable AI, followed by a taxonomy of explainable recommendation approaches from the technical perspective. 

\subsubsection*{\textbf{Differences from Explainable AI}}
The fundamental goal of explainability in AI is to open up the ``black-box'' of the pipelines in AI-related domain, not only to provide trustworthy explanations to users, but also to drive toward more interpretable models. 
As an important type of intelligent decision-making system, the modern recommender system is expected to provide high-quality recommendation results as well as personalized and intuitive explanations with better user engagement, which are important for many practical applications such as e-commerce and social media platforms. Different from broader research area of explainable AI, recommender systems manifest the following characteristics on the requirement of explainability.
\begin{itemize}
    \item \textbf{Personalized Explanation}. Some existing works show that recommender systems should provide different recommendations and explanations to fit different user preferences. \citeauthor{li2021personalized} \cite{li2021personalized} explore the personalized natural language generation in the review summarization and dialog systems. The proposed method addresses personalized generation based on Transformers so as to bootstrap the strength of language modeling to generate high-quality and personalized explanations for recommender systems. \citeauthor{chen2019personalized} \cite{chen2019personalized} propose a novel neural architecture for explainable recommendation over fashion clothes based on both image region-level features and review from user information. The corresponding personalized explanation will be highlighted through some image regions shown to the users. \citeauthor{li2022personalized} \cite{li2022personalized} study user-understandable explanations using prompt learning. By sequential tuning and regarding recommendation process as regularization, it successfully fuses item IDs into the models to generate natural language explanations for recommendations by treating user and item IDs as prompts.
    
    \item \textbf{Interactive Feedback}.
    Providing explanations to users may have downstream impacts, especially in the applications of session recommendation, conversational recommendation, and interactive recommendations compared to general ML tasks where explanations are only associated with one-time predictions. For example, \citeauthor{chen2020towards} \cite{chen2020towards} introduce an explainable recommender system during conversations between the users and agents. {High explanation quality is provided through multi-turn user model conversation.} \citeauthor{omidvar2020interactive} \cite{omidvar2020interactive} conduct Explainable Points-of-Interest (POI) Recommendation as an exploratory process in which users are allowed to keep interaction with the system explanations by expressing their favorite POIs, and the interactions will impact the recommendation process. \citeauthor{wu2021deconfounded} \cite{wu2021deconfounded} provide vision-language explanations by deconfounded learning that conducts pre-training for the vision-language model. In that way, the potential effects of confounders are removed, which will expedite accurate representation training and better explainability. The proposed interactive mechanisms help users to better understand why the system returns particular results and also allow users to effectively return feedback to improve the recommender results.
    
    \item \textbf{Subjective Reaction}. General interpretable machine learning methods populate explanations for understanding the underlying mechanism of how models make predictions so that model developers can better tune or debug the methods. In contrast, explainable recommendations are devised for end users as well, who may have no knowledge about AI at all and will have a subjective understanding of the explanations. As a result, users may react to the explanations differently. \citeauthor{chen2022measuring} \cite{chen2022measuring} claim that recommender system should be considered as a subjective AI task where there is no definite ``right'' or ``wrong'' item to recommend for a user. Instead, it all depends on whether the system can appropriately explain and justify the recommendation.
    Among the methods and evaluation metrics, the satisfaction and trust of users should be carefully considered. The framework in \citeauthor{le2021explainable} \cite{le2021explainable} incorporates both subjective and objective aspect-level quality assumptions and integrates them with recommendation objectives as constraints.
\end{itemize}

\subsubsection*{\textbf{Taxonomy on Explainable Recommendation Approaches}}
To align with the explainable recommendation research in HCI, we propose to categorize explanation generation approaches based on information source of explanations, which are directly related to the design of different methods and inherited display style of output explanations. Furthermore, many existing explainable recommendation methods can be categorized as either model-intrinsic or model-agnostic, as well as global or local explanation. Model-intrinsic or model-agnostic refers to whether the explanation generation process is part of the recommendation step inside the recommendation model or a post-hoc step that generates explanations following the recommendation decision made by models. As a result, one should take caution that recommendation explanations may or may not reflect the internal working mechanism of the decision-making model, such as in post-hoc explanations. 
However, it is important to note that even though some explanations may not be relevant to the model's internal working mechanism, it does not mean that such explanations can be fake explanations. Instead, to assist users' decision making, the explanations shall provide faithful facts about the model or the item, which can be retrieved from the item database \cite{li2021extra}, generated by language models followed by a fact-checking process \cite{xie2023factual}, or created by a counterfactual reasoning process to show how the model output changes responding to certain changes in the input \cite{tan2021counterfactual,tran2021counterfactual}.
The explanations may even talk about the weaknesses of the item on the given feature, as shown by \citet{zhang2014explicit}, \citet{li2020generate} and \citet{musto2021generating}, which explains to users why an item may not be a good match for them, and this helps to increase the trustworthiness of the system from users \cite{musto2021generating}. 
Global or local explanation means whether the explanation is for the general recommendation model as a whole or for an individual recommended item. In the following discussion, we provide information on whether the approach is model-intrinsic or model-agnostic, and populates global or local explanation. By default, the introduced method is model-intrinsic and local, unless specified otherwise.



\begin{itemize}
    \item \textbf{Relevant Users, Items or Features.}
Entity-based explanation refers to a single instance such as similar users, items, or attributes, which can be used as an explanation for the recommended items.
For instance, Explicit Factor Model (EFM) \cite{zhang2014explicit,boost} generates global explainable recommendations via extracting explicit item features as well as user opinions from user reviews in a model-agnostic way. 
TriRank \cite{he2015trirank} identifies important features for explanation from a user-item-feature tri-partie graph.
Tree-enhanced embedding model (TEM) \cite{wang2018tem} learns the decision rules at the first stage, and then an embedding model is designed to incorporate the cross features and generalize the hidden user and item IDs.
Sentiment Utility Logistic Model (SULM) \cite{bauman2017aspect} conducts aspect based recommendation by recommending items with the most valuable aspects from user reviews and meanwhile providing those valuable aspects as explanations. Social explanation \cite{sharma2013social} takes the preference or interactions of the target user's friends to generate explanations for the target user.
Dynamic Explainable Recommender (DER) \cite{chen2019dynamic} proposes the explainable recommendation paradigm dynamically based on the time-aware gated recurrent unit (GRU) modules and represents items based on sentence-level convolutional neural network (CNN).
Extract-Expect-Explain (EX$^3$) \cite{xian2021ex3} generates a set of relevant and similar items as recommendations in the e-commerce area, and a corresponding set of similar attributes is selected to justify the recommended same group of items.
Neural Collaborative Reasoning (NCR) and related works \cite{chen2021neural,shi2020neural,zhu2021faithfully,chen2022graph,zhang2022neuro,xian2020neural} employ explicit neural-symbolic reasoning rules over users, items or attributes to make the recommendation process transparent.

\item {\textbf{Item Descriptions and User Reviews.}}
Textual data widely exists in recommender systems such as item descriptions and user reviews. 
Some explainable recommendation algorithms extract important information from the input text and yield understandable auxiliary sentence justifications as explanations.
For instance, \citeauthor{wang2018reinforcement} \cite{wang2018reinforcement} develop a model-agnostic reinforcement learning framework to generate textual sentence explanations.
\citeauthor{chen2019generate} \cite{chen2019generate} adopt sequence-to-sequence modeling to generate textual sentence explanations.
\citeauthor{li2021personalized} \cite{li2021personalized,li2022personalized} propose personalized transformers and personalized prompt learning to generate fluent and high-coverage explanation sentences, which show significant explanation generation quality.
\citeauthor{pan2022accurate} \cite{pan2022accurate} utilize the review rationalization with rationale generator to extract rationales from reviews to alleviate the effects of spurious correlations when explaining rating predictions. The experimental results demonstrate the improvement of recommendation accuracy as well as novel causal-aware explanations.
\citeauthor{wu2016explaining} \cite{wu2016explaining} develop a succinct additive co-clustering approach on both ratings and reviews to achieve accurate and interpretable recommendations.
\citeauthor{hada2021rexplug} \cite{hada2021rexplug} present an end-to-end framework for explaining recommendations with a sentiment classifier to control the pre-trained language model. The benefit is bypassing the costly training from scratch of the language model which is efficient for generating reviews. 
\citeauthor{wang2018explainable} \cite{wang2018explainable} develop a multi-task learning framework for an explainable recommender system. Specifically, both user preference and content modeling are jointly learned by tensor factorization.

\item {\textbf{Item Images and Videos.}}
This type of data refers to rich multimedia such as images and video, which can provide more intuitive and fascinating demonstrations of items.
\citeauthor{chen2018visually} \cite{chen2018visually} first demonstrate the idea of visually explainable recommendation based on attention neural networks. 
Later, \citeauthor{chen2019personalized} \cite{chen2019personalized} propose a neural architecture for explainable fashion recommendation based on both image region-level features and user review information. 
\citeauthor{cheng2019mmalfm} \cite{cheng2019mmalfm} apply an aspect-aware topic model as the multi-modal over both text reviews and item images so as to better model user preferences and item features from different aspects. The estimation of the aspect importance is also integrated into an aspect-aware latent factor model. The proposed framework alleviates the data sparsity challenge and presents good explainability for recommendation.

\item {\textbf{Logical and Neural-Symbolic Recommendation Rules.}}
Neural-Symbolic rule-based recommender systems can conduct reasoning based on predefined or learned logical rules to make predictions, and the rules can be used to explain the recommendation process.
\citeauthor{shi2020neural} \cite{shi2020neural} unify the power of deep learning and logic reasoning and propose a dynamic neural-symbolic architecture which enables logical reasoning in a differentiable representation space.
Some basic logical operations such as AND, OR, NOT are learned as neural modules based on self-supervised logical regularizer to infer to true or false value of logical expressions.
\citeauthor{chen2021neural} \cite{chen2021neural} further propose Neural Collaborative Reasoning (NCR), which takes explicit reasoning rules for improved transparency in recommendation.
\citeauthor{zhu2021faithfully} \cite{zhu2021faithfully} and \citeauthor{xian2020cafe} \cite{xian2020cafe}  propose knowledge graph (KG) enhanced neural-symbolic reasoning model for recommendation by marrying the interpretability of symbolic rules and the expressiveness of KG embeddings.
Neural-symbolic reasoning is able to guide the path reasoning process to generate faithful explanations, which are demonstrated to be consistent with historic user behaviors and the resulting paths genuinely reflect the decision-making process in KG reasoning. Recently, \citeauthor{zhang2022neuro} \cite{zhang2022neuro} propose attribute-level neural-symbolic reasoning for recommendation, which extracts attribute-level logical rules to further enhance the transparency of the recommendation process.

\item {\textbf{User-, Item-, or User-Item-Related Graphs.}}
Recommendation algorithms can operate over recommendation data structured in graph form such as user-item bipartite graph, item-attribute knowledge graph, and user-interest temporal graph. As a result, mining explanations from the graph structure is important.
\citeauthor{ai2018learning} \cite{ai2018learning} propose a model-agnostic approach to generate path-based explanation from knowledge graph embeddings. Despite boosted explanation performance, the KG paths may not reflect the actual decision-making process of item recommendations, and hence a series of model-intrinsic methods are developed subsequently.
\citeauthor{wang2019explainable} \cite{wang2019explainable} explore path representations by composing the semantics of both entities and relations within the KG. The reasoning procedure from user entities toward item entities over the KG paths will highlight the explanation process of the recommendation.
The pivotal work by \citeauthor{xian2019reinforcement} \cite{xian2019reinforcement} propose a reinforcement knowledge graph reasoning framework, which is able to generate path-based explanations via a policy-guided path reasoning (PGPR) agent to conduct efficient multi-hop reasoning over knowledge graphs. The resulting reasoning paths serve as the explanation since they explicitly expose the multi-step decision-making procedure. 
\citeauthor{xian2020cafe} \cite{xian2020cafe,xian2020neural} further propose neural-symbolic reasoning over knowledge graphs for explainable recommendation,
which learns graph relations as neural-symbolic operators and encodes KG paths as user profiles based on neural-symbolic computation to capture prominent user behaviors in graph reasoning.
\citeauthor{geng2022path} \cite{geng2022path} consider KG paths as sequences of tokens and proposed Path Language Modeling (PLM) for explainable recommendation. Such method includes learning a language model over the knowledge-grounded KG paths for path sequence decoding.
Moreover, the method is capable of conducting explainable recommendation even when the KG structure is sparse and extremely large.
\citeauthor{ma2019jointly} \cite{ma2019jointly} propose a joint learning framework which utilizes knowledge graphs to induce associative explainable rules for item through rule learning.
By extracting rules, the method also shows a better ability to deal with the cold-start recommendation problem.

\item {\textbf{Counterfactuals.}}
Some existing works adopted counterfactual reasoning from causal inference and applied it in recommendation scenarios. For a RS model that takes the input data (e.g., user history, item features.) and makes certain recommendations, counterfactual reasoning looks for what input should be changed and, by how much, to acquire a different prediction. The changed factors could be essential to form the explanation. 
More specifically, \citeauthor{ghazimatin2020prince} \cite{ghazimatin2020prince} propose a searching algorithm on heterogeneous graph to generate model-agnostic explanations, which looks for a minimal set of users' historical actions such that by removing them, the model will recommend different items. \citeauthor{tran2021counterfactual} \cite{tran2021counterfactual} extend influence functions to identify the most relevant training points and deduce a counterfactual set as model-agnostic explanation.
\citeauthor{xu2021learning} \cite{xu2021learning} propose a causality mining algorithm based on input sequence perturbation to extract counterfactual model-agnostic explanations for sequential recommendation. \citeauthor{zhou2021intrinsic} \cite{zhou2021intrinsic} propose counterfactual explanations to improve the explainability of contextualized recommender systems.
\citeauthor{tan2021counterfactual} \cite{tan2021counterfactual} propose a general model-agnostic explainable recommendation framework, Counterfactual Explainable Recommendation (CountER). 
It formulated a machine learning optimization problem to generate simple and effective explanations based on item aspects. Moreover, they also propose a causal evaluation metric to quantitatively evaluated the faithfulness of the generated explanations without access to the ground truth data. Later, as an extension of \cite{tan2021counterfactual}, \cite{tan2022learning} further analyze the relationship between counterfactual and factual reasoning and suggested that by combing both counterfactual and factual reasoning into one framework, the explainable models could generate model-agnostic explanations that are both sufficient and necessary.

\item {\textbf{Multi-round Interactions.}}
\citeauthor{chen2020towards} \cite{chen2020towards} introduce explainable conversational recommendation. In the multi-turn user modeling conversational recommender system, a novel multi-task learning framework that enables tight collaboration between recommendation prediction, explanation generation, as well as user feedback integration is formulated. 
The multi-view feedback integration process integrates user feedback into the recommendation explanations so as to help users understand the model and meanwhile collect user feedback to understand the user interests.
\citeauthor{fu2021hoops} \cite{fu2021hoops} propose a Human-in-the-Loop (HitL) graph reasoning paradigm and created a benchmark dataset to support explainable conversational recommendation research over knowledge graphs. Specifically, the idea is to leverage knowledge graphs to interpret diverse user behaviors. The conversational turns are able to track the human decision-making process while tracing the knowledge graph structures for transparency and explanation generation. 

\end{itemize}

\subsection{Explanations Evaluation in Recommender Systems}

A fundamental problem of explainable recommendation is how to evaluate the explanations \cite{chen2022measuring,zhang2020explainable}.
According to previous works \cite{tintarev2015explaining}, the explanation perspectives can be categorized based on four groups of serving targets:
\begin{itemize}
    \item \textbf{End Users} (satisfaction, trustworthiness). For end users, explanations help them to better understand the features, qualities, and relevance of the items so as to make accurate next-step decisions (click, purchase, etc). In this case, a good explanation is supposed to be informative and useful to end users such that user satisfaction and trustworthiness are maximized. Conversely, offering detailed explanations to them about the computational mechanisms of the recommendation model can also be invaluable.
    \item \textbf{System Developers} (transparency, consistency). For those who develop recommender systems, explanations are mainly used to justify why the predictions are derived from the ``black-box'' models and hence to help developers to debug and scrutinize whether the model works as expected. In this case, the explanation is more measured towards transparency (whether the explanation can reveal model mechanism), and consistency (whether the explanation generation process is consistent with the actual decision-making steps).
    \item \textbf{Content Providers} (effectiveness, efficiency). { On one hand, content providers such as sellers and advertisers care about the effectiveness and efficiency of the additional explanations for the recommended items. The effectiveness of explanation can be measured through common metrics such as click-through rate, conversion rate, etc., while the efficiency depicts whether displaying explanations can expedite the end-user decision-making process. On the other hand, providing content providers with tailored explanations about the recommendation mechanisms of their products can be of immense value to them, while such evaluation remains an open problem.}
    \item \textbf{Regulators} (scrutability). Regulators can leverage explanations to examine whether the recommender systems properly use users' sensitive personal data. Therefore, the measuring mechanism is similar to those for system developers where the resulting explanations are supposed to be transparent.
\end{itemize}


The explanation evaluation approaches can be generally divided into the following three categories, each of which faces a trade-off between result reliability and evaluation cost.
\begin{itemize}
    \item \textbf{Offline Evaluation}. It refers to using offline datasets and quantitative metrics to evaluate the quality of explanations generated by different explainable recommendation approaches. However, unlike other objective AI tasks such as image classification, where data labels are usually objective facts about the images and do not change with human's subjective preferences, explanations tend to be more subjective and there is usually no absolute ground-truth explanation for a recommendation \cite{herlocker2000explaining}. As a result, many papers leverage the user reviews which represent comprehensive user preferences as the ground-truth explanation \cite{chen2022measuring}. Recent research also tries to develop explainable recommendation datasets with human labeled ground-truth explanations to help the research of model development and evaluation \cite{chen2023reasoner}. Depending on whether or not a ground-truth explanation is available and the forms of output explanations such as natural language explanations and instance-based explanations, the metrics can vary under different settings \cite{chen2022measuring}. When a ground-truth explanation is available, we can use standard metrics to evaluate the extent that model generated explanation matches with the ground-truth explanation, such as precision, recall, and coverage for feature- or instance-based explanations \cite{zhang2014explicit,chen2019personalized,li2020generate}, BLEU and ROUGE for natural language explanations \cite{li2021personalized}, as well as NDCG, MRR and Hit Ratio for explanation ranking methods \cite{li2021extra}.
    When ground-truth explanation is not available, we can use Probability of Sufficiency (PS) and Probability of Necessity (PN) to evaluate the sufficiency and necessity of explanations, such as evaluating counterfactual explanations \cite{tan2021counterfactual,tan2022learning}.
    There is almost no cost to set up such evaluation process since it is equivalent to measuring model prediction on offline datasets. However, the correlation between these offline metrics and actual user comprehension and utilization of recommendation explanations remains unclear and warrants further research.
    \item \textbf{User Study and Online Evaluation}. This set of approaches
    recruit volunteers or select (a subset of) real users from commercial systems and manually evaluate the explanation quality through a simulated or real-world environment. The evaluation process could be either active (actively ask users some survey questions to evaluate the explanations) or passive (passively track and record some user reactions when interacting with explanations).
    For example, \citeauthor{zhang2014explicit} \cite{zhang2014explicit} conduct A/B testing in a simulated environment over online users to evaluate how explanations influence users' click behaviors, while \citeauthor{xian2021ex3} \cite{xian2021ex3} conduct large-scale A/B testing in Amazon real-world commercial system to examine the effectiveness of increasing conversion and revenue when providing explanations in e-commerce recommendations. 
    Online evaluation can better reflect users' reactions to explanations in real environments, but the experiment cost is also much higher and sometimes the evaluation environment is not accessible to researchers.
\end{itemize}

From existing research works, offline evaluation and survey-based user studies are widely adopted since they are doable and easy to set up, while online evaluation is less frequent since it requires access to the real-world system. We summarize common metrics adopted in these evaluation approaches as follows, which mainly depend on explanation forms.
For natural language explanations, common metrics are BLEU and ROUGE \cite{sun2020dual,li2020generate,li2021personalized,chen2019co,hada2021rexplug,yang2021explanation}. Besides, some outstanding work have extended and proposed several novel metrics in explanation evaluation. For example,
\citeauthor{li2020generate} \cite{li2020generate,li2021personalized} not only evaluate the generated sentence explanations but also evaluate how well the sentences really explain the recommendations based on newly designed metrics such as Unique Sentence Ratio (USR), Feature Matching Ratio (FMR), Feature Coverage Ratio (FCR) and Feature Diversity (DIV), which evaluates the uniqueness and personalization of the sentence explanations.
For matching or ranking-based explanation (KG path, features, etc.), the common evaluation metrics are NDCG, Precision, recall and coverage
\cite{tai2021user,ren2017social,yang2020meta,symeonidis2009moviexplain,li2021extra}. For the evaluation of counterfactual explanation, common metrics are Average Treatment Effect (ATE), replacement, Probability of Sufficiency (PS) and Probability of Necessity (PN) \cite{chen2020try,tran2021counterfactual,kaffes2021model,tan2021counterfactual,tan2022learning}. Some other evaluation metrics include Perplexity, Mean Explanation Precision (MEP), Mean Explanation Recall (MER) as well as influence \cite{lu2018like,abdollahi2017using,liu2019in2rec,peake2018explanation}.

\subsection{Relationship with other Trustworthy Recommendation Perspectives}
\label{sec:relationship_explainability}
The importance of explainability in trustworthy recommendation systems cannot be overstated, as it has a significant impact on the users' trust over the items and the platform. When users do not receive clear and informative explanations for a recommended item to aid their decision making, they may become frustrated or lose faith in the system, which may further prompt them to explore other recommendation sources or even leave the platform entirely.
Similarly, items may suffer from decreased trust, as users may be less likely to click the ads or purchase the items if they cannot understand why they have been recommended. Such effects can negatively impact the reputation of the platform, as well as the overall revenue of the platform.
Overall, it is essential for designers and developers of recommendation systems to prioritize transparency and provide clear explanations to users about why recommendations are being made.
Furthermore, the importance of explainability in RS extends beyond enhancing user experience---it is closely related to other RS trustworthiness perspectives, which we discuss in the following.


\begin{itemize}
    \item \textbf{Fairness-aware Explainability} aims to generate explanations that are fair. The ultimate goal of explanation is to help users understand the model decisions and thus to gain trust. However, research works find that the recommendation explanations may be vulnerable to bias and unfairness \cite{fu2020fairness, zhao2022fairness}, i.e., the explanation quality for different user groups could be very different. More specifically, the explanation system may tend to generate stereotyped explanations, i.e., generating some type of explanations for one group of users while generating another type of explanations for another group of users so as to attract users to click the recommended items, regardless of the real underlying mechanism that leads to the model decisions \cite{zhao2022fairness}. For example, the system may explain a coat as ``fashionable'' for a young consumer while explaining the same coat as ``warm'' for an elderly consumer due to the stereotype patterns that the system learns from data, even though the elderly consumer actually also cares about the fashion of clothes.
    If this is the case, explainable AI may completely violate its original goal---instead of gaining trust, it may actually lead to total distrust if users realize that the explanations are biased and they are manipulated or even cheated by the explanations to click recommendations. Therefore, generating fairness-aware explanations is an important consideration. To date, seldom progress has been testified in the field of unbiased and fairness-aware explanations, either identifying the underlying reason for model disparity in recommendation \cite{ge2022explainable} or exploring the trade-off between explainability and fairness/unbiasedness in trustworthy recommender system. To solve the problem, extensive efforts are needed to guarantee that explanations are faithful, unbiased, and fair.
    
    \item \textbf{Privacy-aware Explainability} aims to protect user privacy when generating explanations. Providing explanations to users can help them better understand system outputs and increase user trust in the system. However, providing explanations may also reveal users' information in unwanted ways, potentially compromising user privacy. For example, a frequently used explanation in social recommendation is neighbour-user explanation, which tells the target user that the reason a certain item is recommended is because his or her friend(s) clicked this item. However, this may unwillingly leak the private information of these friend(s). As a result, how to anonymize such information and meanwhile generating informative explanation is an important problem \cite{georgara2022privacy,brunotte2021can}, and it is crucial to develop privacy-aware explanation techniques to ensure that the system can provide explanations while also protecting user privacy. Further research in this area is needed on developing techniques that strike a balance between providing helpful explanations and maintaining user privacy.

    \item \textbf{Robust Explainability} aims to increase the recommender system's robustness to attacks while providing explanations. Since providing explanations may reveal the recommender system's internal working mechanism to outside parties, if the explanations are leveraged by adversary entities, then it may expose the system to attacks by helping the adversary entities to create more targeted attacking methods. Counterfactual explanation \cite{ghazimatin2020prince,tran2021counterfactual,kaffes2021model,tan2021counterfactual} is especially vulnerable to malicious usage, because it directly tells users what interventions can change the recommendation result in what ways \cite{slack2021counterfactual,aivodji2020model}. Therefore, it is crucial to develop robust explainable recommendation systems.

    \item \textbf{Controllable Explainability} aims to improve the user controllability in the explanation generation process from different perspectives, such as the level of detail and the level of expertise. For the level of detail, controllable explanation helps to explain the property of an item over the feature(s) specified by the user, so that users can easily understand the key properties of the item on the feature(s) that the user cares about, because users usually want to know more details about an item before deciding whether to go with the recommendation or not. Since practical recommender systems and explanations usually involve a large amount of features, it can be difficult and time-consuming for users to go through each and every feature of an item. Rather, users may be interested in the property of an item over only a few features, and users' preference on features can be highly personalized. One example is natural language explanation generation \cite{li2020towards}. For instance, sometimes the user may care about the quality of the item and thus hope to know more about its quality feature, while other times the user may be interested in the brand feature and thus hope to know more about its brand. As a result, such explanations can provide users with the flexibility of choosing the target feature(s) so as to describe the item on a subset of perspectives \cite{li2020generate,xie2023factual,li2021personalized}, leading to different levels of detail for the explanations.
    Users may also control the expertise level of the explanations, i.e., if users are experts on recommender systems, then they may want complex and complete explanations; otherwise, they may want simple or partial explanations, allowing level-of-expertise controllability.
    In addition, they might desire greater control over the length and sentiment of the explanation, which requires joint consideration of controllable generation \cite{zhang2022survey} and explainable recommendation \cite{zhang2020explainable}.
    However, it is important to note that even though the system may grant users greater controllability of the explanations, it does not mean that users can manipulate the system to generate fake explanations. Instead, to assist users' decision making, the explanations shall provide faithful facts about the item on the provided feature, which can be retrieved from the item database \cite{li2021extra} or generated by language models followed by a fact-checking process \cite{xie2023factual}. 
    Moreover, implementing user-friendly and unambiguous controllable explanation interfaces within real-world systems is also crucial to users. The primary objective is to safeguard against any potential misleading of users by the system, thus reinforcing trust and faithfulness in the interaction between human users and recommender systems \cite{tsai2019designing}.


\end{itemize}


%% file: sections/fairness.tex
\section{Fairness}\label{sec: fairness}
Recommender Systems have been considered as ``benevolent'' systems for a long time, which assist users (e.g., by helping them find relevant items) and create value for businesses (e.g., higher sales or increased customer retention) \cite{10.1145/3370082}. However, in the most recent years, considerable concerns from both academia and industry have been raised regarding the issue of fairness in recommendation ~\cite{li2022fairness}.
Several studies argue that RS may be vulnerable to unfairness in several aspects, which may result in detrimental consequences for underrepresented or disadvantaged groups~\cite{Geyik2019,singh2018fairness,li2021cikm,li2021user}. For example, in e-commerce systems, RS may promote items that mainly maximize the profit of certain producers \cite{ge2021towards}, or in online job marketplaces, RS may lead to racial or gender discrimination by disproportionately recommending low-payment jobs to certain user groups \cite{Geyik2019}.
Therefore, to improve the satisfaction of different stakeholders in RS \cite{abdollahpouri2019multi}, it is important to study fairness in recommendation and build trustworthy and responsible systems.

\subsection{Source of Unfairness in Recommendation}
Bias and discrimination are two common concepts related to the unfairness issue in machine learning and unfairness stems from both of these factors \cite{mehrabi2021survey,li2022fairness}.
In recommender systems, there are two main types of biases: bias in data, and bias in algorithm \cite{li2022fairness}.
The biases in data may come from the processes of data generation, data collection, or data storage. For example, data bias may be caused by the specifics of the data collection process such as when a biased sampling strategy is applied. 
When training on the biased data, the recommendation models are highly likely to learn those over-represented groups, promote them in the ranked results, and potentially result in systematic discrimination and reduced visibility for disadvantaged user/item groups such as under-representing the minorities, certain racial groups, or gender stereotypes \cite{li2021user,li2021towards}.
Another source of unfairness may lie in the recommendation model itself. For example, recommendation models may further reinforce existing biases or existing skewed distributions in the underlying data.
A well-known recommendation issue caused by such bias is popularity bias, where the popular items with more user interactions will be recommended more frequently and get more exposure opportunity than those less popular but equally or even more relevant ones \cite{ge2021towards}.
Apart from bias, discrimination is caused by intentional or unintentional human prejudices and stereotyping with regard to sensitive attributes (e.g., race, gender, and religion, etc.) \cite{mehrabi2021survey}. It is worth noting that bias and discrimination are not the only reasons of unfairness. For example, research has shown that some fairness definitions cannot be satisfied simultaneously \cite{chouldechova2017fair,kleinberg2017inherent,pleiss2017fairness}, as a result, the violation of one fairness definition may be caused by ensuring another fairness definition \cite{li2022fairness}.

\begin{figure}[t!]
  \centering
  \includegraphics[width=0.85\textwidth]{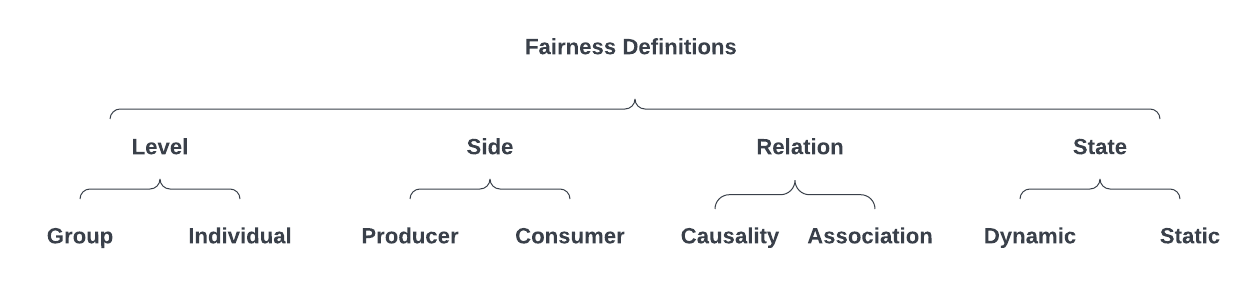}
  \caption{Different dimensions to define fairness in recommendation.}
  \label{fig:fairness_definition}
\end{figure}

\subsection{Definitions of Fairness in Recommendation}
{Fairness is a complex and multifaceted concept that can be defined on various different perspectives. An abundance of definitions has been explored in many research areas and tasks. Considering fairness in the recommendation scenario makes the landscape even more convoluted.
To clearly introduce the progress of fairness in recommender systems, we start from introducing its definitions.
One of the closest definition of fairness in recommendation comes from \cite{ekstrand2022fairness}, which asserts, 

\begin{displayquote}
\textit{When we refer to fairness, we are talking about the ways a system treats people, or groups of people, in a way that is considered ``unfair'' by some moral, legal, or ethical standard. This is typically through effects or impacts that are not experienced in an equitable way, but can sometimes arise through the system's internal operation or representations.}
\end{displayquote}

With a clear understanding of the definition of fairness in recommendation scenario, we now categorize fairness into different classes from different perspectives. Particularly, we distinguish between individual and group fairness, consumer and producer fairness, associative and causal fairness, as well as static and dynamic fairness (Figure \ref{fig:fairness_definition}). }

Fairness aims at ``the absence of any prejudice or favoritism towards an individual or a group based on their intrinsic or acquired traits in the context of decision-making'' \cite{lan2021proof,mehrabi2021survey}.
Therefore, in the first place, fairness is defined on group-level and individual-level.

\begin{itemize}
    \item \textbf{Group Fairness} is the idea that the average treatments should be the same across groups defined by certain attributes \cite{varshney2019trustworthy}.
    Such attributes are called protected (or sensitive) attributes, which often include gender, religion, age, sexual orientation and race, among others.
    Based on the above definitions, many variants are derived, such as Equal Opportunity, which requires that the true positive rate is the same across different groups \cite{hardt2016equality}, Equalized Odds, which requires that different groups should have the same true positive rate and false positive rate \cite{hardt2016equality,zafar2017fairness}, as well as Demographic Parity, which requires that each group should have the same likelihood to be classified as positive \cite{calders2009building}. In recommendation scenarios, \citeauthor{li2021user} \cite{li2021user} consider user-oriented group fairness. Specifically, they divide users into active and inactive groups based on the number of user interactions in the training data, and require that different user groups should receive similar recommendation quality such as F1 and NDCG.
    
    \item \textbf{Individual Fairness} is the idea that individuals similar in their features should receive similar model predictions, namely, similar individuals should be treated similarly \cite{dwork2012fairness}. 
    For example, \citeauthor{lin2017fairness} \cite{lin2017fairness} consider each individual user's utility of a recommended item as the relevance of the item to the user, and then considered fairness as the imbalance between users' utilities.
    \citeauthor{li2021towards} \cite{li2021towards} study counterfactual fairness in recommendation, which is a type of individual fairness and requires that the recommendation results for each user should be same or similar in both the factual and the counterfactual world. 
    
\end{itemize}

Considering that the fairness demands in recommender systems may come from different stakeholders, fairness in recommendation is also divided into user (consumer)-side fairness and item (producer)-side fairness.
Besides, there are also cases in which a system may require fairness for both consumers and producers, when, for instance, both users and items belong to protected groups.

\begin{itemize}
    \item \textbf{User (consumer)-side Fairness} studies the disparate impact of recommendation on protected classes of consumers \cite{zehlike2021fairness}. The protected classes can be objective features such as race and gender, while it can also be subjectively assigned features. For example, \citeauthor{Yao2017} \cite{Yao2017} investigate gender-based inequalities in collaborative filtering recommender systems, while \citeauthor{li2021user} \cite{li2021user} group users based on their interaction frequency with the recommender and found that the active users accounting for a small proportion of the users (5\%) enjoy much higher recommendation quality than others (95\%). Even though most of the existing works focus on group-level user-side fairness, it can also be defined from individual-level. For example, \citeauthor{li2021towards} \cite{li2021towards} leverage counterfactual fairness which requires that the recommendation results for each user are unchanged in the counterfactual world where the user's sensitive features are flipped, and each individual user can specify their sensitive features by themselves.

    \item \textbf{Item (producer)-side Fairness} considers fairness for the items and the item producers in the recommender system, which ensures market fairness and avoids monopoly domination or Matthew's Effect \cite{zehlike2021fairness}. For example, \citeauthor{ge2021towards} \cite{ge2021towards} and \citeauthor{abdollahpouri2017controlling} \cite{abdollahpouri2019unfairness, abdollahpouri2017controlling} focus on item popularity bias since popular items (i.e. those frequently rated, clicked or purchased items) get disproportionately more exposure while less popular ones are under-recommended.
    Moreover, some researchers explore producer fairness based on the sensitive attribute of the item producers such as their gender  \cite{gharahighehi2021fair,ferraro2019music,boratto2021interplay,kirnap2021estimation,shakespeare2020exploring,gomez2021winner}, which is similar to the exploration of consumer fairness based on sensitive features.
\end{itemize}


The research community has studied fairness in machine learning by developing association-based (or correlation-based) fairness notions for a long time and most of the existing works about fairness in recommendations consider the association-based fairness notions. 
However, recently, some pioneering works have found that fairness cannot be well assessed only based on association notions \cite{khademi2019fairness,kusner2017counterfactual,zhang2018equality,zhang2018fairness,li2021towards}, since such fairness definitions cannot reason about the underlying causal mechanism that leads to unfairness. As a result, we also introduce associative and causal fairness definitions.

\begin{itemize}
    \item \textbf{Associative Fairness} is also known as correlation-based/statistical fairness, which measures the statistical discrepancy between individuals or sub-populations, such as Equal Opportunity \cite{hardt2016equality}, Equalized Odds \cite{berk2021fairness} and Demographic Parity \cite{dwork2012fairness,zafar2019fairness}.
    One weakness of associative fairness is that they cannot detect discrimination in presence of statistical anomalies such as Simpson's paradox, namely, the statistical conclusions drawn from the sub-populations could differ from that from the whole population \cite{khademi2019fairness,kusner2017counterfactual,zhang2018equality}.
    
    \item \textbf{Causal Fairness} is not only based on data but also considers additional prior knowledge about the structure of the world in the form of a causal model. Besides, it is important to explore the causal relationship between the sensitive attributes and model output rather than just the associative relationship. 
    {
    \citet{wang2022causal} utilize causal relationships to address fairness concerns in the domain of recommender systems, particularly in the context of multi-behavior recommendation scenarios. They employ two causal interventions to tackle the issue of unfairness, and additionally use back-door adjustment to estimate the causal effect of their interventions.
    }
\end{itemize}

Fairness in machine learning has predominantly been studied in static classification settings without concerning how decisions change the data over time.
However, the fairness requirements in recommendation need to consider the dynamic nature of the systems since many features are changing over time such as user preference and item popularity.
\begin{itemize}
    \item \textbf{Static Fairness} provides a one-time fairness solution based on fairness-constrained optimization, which focuses on the fairness implications of decisions made in a static or one-shot context. Most existing fairness recommendation research put themselves in the static setting \cite{li2022fairness}.

    \item \textbf{Dynamic Fairness} considers the dynamic factors in the environment and learns a fairness strategy that accommodates such dynamics. For example, \citeauthor{ge2021towards} \cite{ge2021towards} study the dynamic fairness of item exposure in recommender systems. The items are separated into popular and long-tail groups based on the number of exposure in training data. The intuition of the work is that the item popularity may change during the recommendation process based on the recommendation strategy and user feedback, causing the underlying group labels to change over time, i.e., an item that was once unpopular may now become popular, and vice versa. To solve the problem, the authors formulate the problem as a Constrained Markov Decision Process (CMDP) with fairness constraint of item exposure change over time, and use Constraint Policy Optimization (CPO) to solve the formulated problem.
\end{itemize}

\begin{figure}[t!]
  \centering
  \includegraphics[width=1\textwidth]{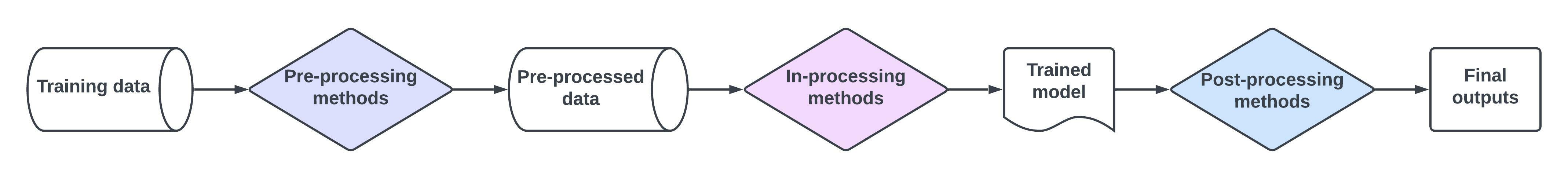}
  \vspace{-15pt}
  \caption{Three types of fair recommendation methods in different parts of the recommendation pipeline.}
  \vspace{-10pt}
  \label{fig:fairness_methods}
\end{figure}

\subsection{Methods for Fairness in Recommendation}
Existing research on fairness in recommendation mostly focuses on three areas: 1) fairness quantification, which develops quantitative metrics to measure algorithm fairness under various fairness definitions, 2) fair recommendation modeling, which develops algorithms or models to improve fairness of the outputs,
and 3) fairness diagnostics, which develops explainable fairness methods to identify the reason of model unfairness so as to explain why a model is fair or unfair.

\subsubsection*{\bf Fairness Quantification} The study of fairness quantification aims to develop and explore quantitative metrics that can measure algorithmic disparities in ranking or recommendation \cite{fu2020fairness,gao2021fair,li2021user}. This line of research has resulted in the identification of various forms of unfairness in recommendations, such as unfairness based on sensitive features like gender and age \cite{chen2018investigating,Yao2017}, fairness based on popular and unpopular items \cite{ge2021towards,abdollahpouri2017controlling,abdollahpouri2019unfairness}, and fairness in recommendation quality for users \cite{fu2020fairness,li2021user}.
For instance, \cite{fu2020fairness,li2021user} propose and investigate a measure of recommendation quality unfairness between active and inactive users. This is one of the examples of how the research on fairness quantification has contributed to identifying and measuring algorithmic disparities in recommendation systems.


\subsubsection*{\bf Fair Recommendation Models} Fair recommendation models focus on providing fair recommendation results based on certain fairness definitions, which can be roughly divided into three categories: pre-processing methods, in-processing methods and post-processing methods \cite{li2021cikm,li2022fairness}. Figure \ref{fig:fairness_methods} illustrates the differences between them. { Every type of method possesses its own set of benefits and drawbacks. It is crucial to be meticulous in our selection or development of methods for enhancing fairness, as the decision-making process may not be purely technical, it requires consideration of social and legal contexts as well.}

\begin{itemize}
    \item \textbf{Pre-processing methods} usually aim to minimize the bias in the source data before training the model.
    Therefore, they can be adopted when we have access to the data and usually they do not explicitly include fairness metrics that are defined over model outputs.
    Representative pre-processing methods include fairness-aware sampling techniques in the data collection process to cover items of all groups, balancing techniques to increase the coverage of minority groups, and repairing techniques to ensure label correctness \cite{10.1145/3404835.3462807}.
    For example, \citeauthor{lahoti2019ifair} \cite{lahoti2019ifair} introduce a method for mapping user records into low-rank representations that reconcile individual fairness through pre-processing. The method operates on an individual fairness objective to learn fair representations of training data points. The proposed method aims to transform an input feature vector into a fairer representation such that the individuals who are indistinguishable on their non-sensitive features in data should also be indistinguishable in their vector representations under a given distance function. { Pre-processing methods have the benefit of allowing any downstream algorithm to be trained with the transformed data. However, pre-processing methods might experience an unforeseen loss in model performance, and can only be used if modifying the training data is both technically and legally permissible.}

    {\item \textbf{In-processing methods} aim to eliminate the bias during the model training process by modifying existing models or introducing new models.
    A general approach is to encode the fairness requirement as part of the objective function, such as a regularization term \cite{abdollahpouri2017controlling,beutel2019fairness,ge2021towards,li2021user} or adversarial term \cite{wu2021learning,li2021personalized,wu2022selective}, which measures the degree of unfairness that the model must minimize in addition to the minimization of the original loss function. This approach also seeks to find a balanced trade-off between recommendation accuracy and fairness during training,
    such as learning the Pareto frontier between fairness and utility so as to find the best balance between the two \cite{ge2021towards,ge2022toward}.
    In-processing methods usually offer greater flexibility to deal with the accuracy-fairness trade-off problem compared with pre-processing and post-processing methods. However, in-processing methods may result in a non-convex optimization problem and cannot ensure optimality.
}
    \item \textbf{Post-processing methods} aim to modify the presentation of the already produced outputs to improve fairness through techniques such as re-ranking by linear programming \cite{li2021user,singh2018fairness,yang2021maximizing} or multi-armed bandit \cite{celis2019controlling}.
    For example, \citeauthor{zehlike2017fa} \cite{zehlike2017fa} propose a fair ranking algorithm---FA*IR to ensure that the number of protected candidates does not fall far below a required minimum percentage $p$ at any point in the ranking. The method formulates this fairness as a statistical significance test of whether a ranking was likely to be produced by a Bernoulli process. \citeauthor{li2021user} \cite{li2021user} propose a fairness re-ranking method which re-ranks the recommendation list of each user to guarantee fair recommendation quality for advantaged and disadvantaged users. {Similar to pre-processing methods, post-processing methods can also provide the model-agnostic flexibility. The post-processing methods, however, are inapplicable if sensitive feature information is not accessible at the decision time.} 
\end{itemize}

\subsubsection*{\bf Fairness Diagnostics} It focuses on answering a more fundamental question: what are the reasons that cause model unfairness?
In general AI, there have been several pioneering works trying to derive explanations for model fairness \cite{begley2020explainability,pan2021explaining}.
For example, \citeauthor{begley2020explainability} \cite{begley2020explainability} leverage Shapley value \cite{shapley201617} to attribute the feature contributions to the model disparity so as to generate explanations.
The proposed method estimates the sum of individual contributions from input features and thus understands which feature contributes the most to the model disparity \cite{begley2020explainability}.
Though this type of methods can provide explanations to the model disparity, they are not suitable for recommender systems due to the large item/user feature space in recommendation.
To solve the problem, \citeauthor{ge2022explainable} \cite{ge2022explainable} design a learning-based counterfactual reasoning method to discover critical features that significantly influence the fairness-utility trade-off and use them as fairness explanations for black-box feature-aware recommendation systems.


\subsection{Relationship with other Trustworthy Recommendation Perspectives}
{
The impact of fairness in recommendation systems can be far-reaching, affecting consumers, items, and the platform as a whole. Unfairness in the recommendation process can result in negative outcomes, such as discrimination, inequality, and decreased trust in the system.
Unfairness in recommendation systems can have several negative impacts on consumers. For example, if the system is biased towards a particular group, it may limit the exposure of certain products or services to users outside of that group. This can result in limited choices for consumers, reducing their ability to make informed decisions. In turn, this can decrease consumer satisfaction and trust in the system.
Not only users but also items can also be affected by unfairness in recommendation systems. For example, biases in the recommendation process can lead to certain items being unfairly promoted or ignored. This can result in an unequal distribution of exposure and sales for different items, leading to decreased revenue for some items and increased revenue for others. 
Finally, unfairness in recommendation systems can negatively impact the platform itself. If the system is perceived as unfair or biased, it can lead to decreased trust and satisfaction among users. This, in turn, can result in consumers and producers leaving the platform, reducing engagement and revenue. Moreover, if the platform is found to be promoting discriminatory or biased recommendations, it may face legal actions, further damaging its reputation and financial standing.
Given the potential negative impacts of unfairness in recommendation systems, it is essential to design and evaluate these systems for fairness. Furthermore, the influence of fairness can be extended to other trustworthy perspectives in recommender systems, including:
}

\begin{itemize}
    
    

{
    \item \textbf{Explainable Fairness} aims to generate explanations so as to help users, system designers or policy makers to understand why the recommender system is fair or unfair, which has gained significance in the field of trustworthy recommendation \cite{ge2022explainable,ghosh2022faircanary,zhou2022towards}. This is because the features that lead to unfairness problems in recommender systems may not be those obvious features such as gender, race or age, which is due to recommendation models ability of collaborative learning from users' interaction data. Actually, most recommendation models do not directly use such sensitive features at all, but still have fairness problems due to collaborative learning and bias propagation. As a result, it is usually very difficult for users, system designers or policy makers to manually decide which feature(s) lead to unfairness problems \cite{ge2022explainable}. This pertains to the ability of algorithms to offer users and system designers a rationale for why a particular model output is regarded as fair or unfair. The explainable fairness perspective seeks to enhance the transparency, comprehensibility, and interpretability of the fairness related decision-making process of the system. By providing a lucid demonstration of the functioning of the recommender system, users and designers can develop greater trust in the fairness of the system \cite{ge2022explainable}.
}
    \item \textbf{Privacy-aware Fairness} aims to guarantee system fairness in a privacy-preserving manner. This poses unique challenges in recommender systems for two reasons: 1) the system may need access to users' sensitive features so as to increase the fair treatment for disadvantaged users; however, 2) users may be unwilling to share their sensitive features with the system for privacy concerns. Sometimes, the reason that users are unwilling to share their sensitive features is exactly because they worry about being unfairly treated by the system, even though the system aims to increase their fairness, which causes a dilemma for recommender systems. One approach to solving the privacy-aware fairness problem is federated fairness \cite{liu2022fairness,maeng2022towards,zhu2022cali3f}, which adopts the federated learning framework for fairness so that users can keep their sensitive information on their personal device and do not have to explicitly share such information with the model. 

{
    \item \textbf{Robust Fairness} aims to develop fairness-aware recommendation algorithms that can withstand intentional or malicious attacks in the recommendation process \cite{solans2021poisoning,zhang2022pipattack,deldjoo2021survey,anelli2021adversarial}. Robust fairness ensures that the system remains fair even when it is targeted by adversaries that aim to make the system unfair. This is particularly important in ensuring that the system is capable of consistently providing unbiased recommendations. Besides, to evaluate the effectiveness of robust fairness techniques, it is essential to use appropriate evaluation metrics, such as worst-case fairness violations or average-case accuracy. Additionally, it is crucial to evaluate the trade-off between robustness and accuracy since increasing robustness often comes at the cost of reduced accuracy \cite{zhang2019theoretically}.

    \item \textbf{Controllable Fairness} is to offer users autonomy over the features they consider essential for fair treatment. It is not about turning the system's fairness on or off, but rather allowing users to prioritize which aspects of fairness they value most, based on their unique life experiences and perspectives. It has been theoretically proven that some notions of fairness are inherently contradictory and cannot be achieved globally at the same time \cite{kleinberg2017inherent}. Therefore, constructing a controllable fairness system, where users can select the kinds of fairness they care about most, becomes a feasible solution. There have already been several pioneering works that focus on a similar idea, such as \cite{wu2022selective,li2021towards}, which allow users to specific the features that they are most sensitive about, and the system can ensure counterfactual fairness on these features when delivering recommendations for them. Meanwhile, since users' personal biases may be a source of unfairness, it could also be a risk if users are allowed to control the fairness of recommender systems. As a result, controllable fairness implementation needs to take extreme care so that users' personal biases do not abruptly interfere with the fairness mechanism of the system. For example, the system can implement bounds or limits to ensure that user-driven fairness controls do not lead to grossly unjust or biased outcomes. In this case, while users might be allowed to prioritize certain fairness aspects, they should not be able to entirely disable fairness mechanisms, especially in contexts where it is ethically and legally mandated. In addition, the system needs to maintain proper fairness-utility relationship when users adjust their fairness demands, which requires the system to adjust its recommendation policy for this user in a dynamic way in light of the user's dynamic preferences on fairness and utility. Finally, the interface that allows users to control fairness can also play an educational role. By presenting users with choices and possibly the implications of those choices, the system can raise user awareness about fairness. Over time, this could foster a more informed and responsible user base.}

\end{itemize}


%% file: sections/privacy.tex
\section{Privacy}\label{sec: privacy}

With the growing concerns about the machine learning methods that gather and analyze personal data, the ethical demand for data privacy has been formally recognized in terms of mandatory regulations and laws \cite{voigt2017gdpr,pardau2018california}.
As a consequence, the research of privacy-preserving machine learning has witnessed substantial development in recent years \cite{liu2021machine}.
It is believed that a more trustworthy web service would provide privacy-protected solutions that avoid unwanted exposure of information for any participants of the system.

In both the recommender system and the general machine learning field, there exist several definitions of privacy \cite{kang1997information,solove2005taxonomy,smith2011information,aghasian2018user}, and in most cases, they share the same ingredients:

\begin{itemize}
    \item \textbf{Private information}: the critical or valuable information that needs restriction of access. For example, user identity and sensitive user attributes such as gender, age, and address.
    \item \textbf{Ownership}: only the authorized entities can access and control the corresponding private information, where the entity may refer to a user or even the platform itself.
    \item \textbf{Threat}: malicious entities (inside or outside the system) aiming to get access to or manipulate private information. 
    Note that such entities may utilize auxiliary \textit{public information} to engage its infiltration or attacks.
    \item \textbf{Goal of privacy protection}: to maintain the \textit{ownership} of the \textit{private information} and find countermeasures for the \textit{threats}.
\end{itemize} 

In this section, we adopt these terminologies and discuss the privacy problems in the field of recommender systems (RS).
We first explain the privacy demands of different types of ownership in RS in Section \ref{sec: privacy_ownership}, and then list the major privacy threats and challenges in Section \ref{sec: privacy_threats}.
Section \ref{sec: privacy_protection} illustrates several major privacy-protection techniques.
And finally, we enumerate several open questions in Section \ref{sec: privacy_open_problems}.
For existing surveys, we consider \cite{friedman2015privacy} as the backbone of our taxonomies, then include and extend the ideas from several recent surveys \cite{aghasian2018user,wang2018toward,himeur2022latest}.

\subsection{Ownership of Private Information}
\label{sec: privacy_ownership}

While the term ``privacy'' is used in a variety of scenarios on the Web, privacy problems in RS are typically related to two types of entities: users/customers and the recommendation platform itself.
Each type of entity corresponds to a specific type of ownership demand and thus faces different privacy risks.

\textbf{User Privacy}: To provide accurate personalized predictions, recommender systems heavily depend on rich user digital traces.
{
This may involve the collection of sensitive or critical user information and user preferences, such as browsing history, both of which may potentially threaten users' privacy.}
Recent debates and regulations \cite{voigt2017gdpr} on user's data privacy brought up concerns about platforms gathering sensitive user information such as gender, race, age, location, sexual orientation, and contact information, and potentially using it in unlawful or untrustworthy ways, such as targeted advertising and scamming.
In RS, this is a natural conflict since most recommendation solutions are personalized and require such information to construct accurate user profiles.
Though some research found that users may become less concerned about privacy leakage to trusted services \cite{crocco2020s}, the trade-off between recommendation personalization and user privacy has been one of the critical design challenges for modern recommender systems \cite{awad2006personalization,li2012willing}.

\textbf{Platform Privacy}: Even if the recommendation platform rightfully collects and utilizes the data from users, there are still privacy risks if outside attackers infiltrate the system or reveal critical system information such as the log data and model parameters.
On one hand, when the RS does not provide sufficient system security and data anonymization, adversaries may recover the user's data and system information through hacking or inference \cite{calandrino2011you}.
On the other hand, one can pretend to be a user or a trusted third-party, and interact with the RS in order to influence the decision system.
For example, a ``fake user'' can inject disguised data that is specifically designed to trick the RS, so that the resulting recommendation model learns to favor certain items or business owners and suppress the others \cite{fang2018poisoning}.

In this section, we address the difference between these two types of ownership from the perspective of trust management: while the user privacy problem mainly concerns the users' trust towards the RS platform, the platform privacy discusses the system's trust towards its users and external entities.
In general, the RS should show both its legitimacy in collecting users' data and its ability to protect its data and model against privacy attacks.
For the following sections, we first discuss existing research works in terms of the privacy threats and then list several open questions in the field according to this taxonomy.

\begin{figure}[t!]
    \centering
    \includegraphics[width=0.9\textwidth]{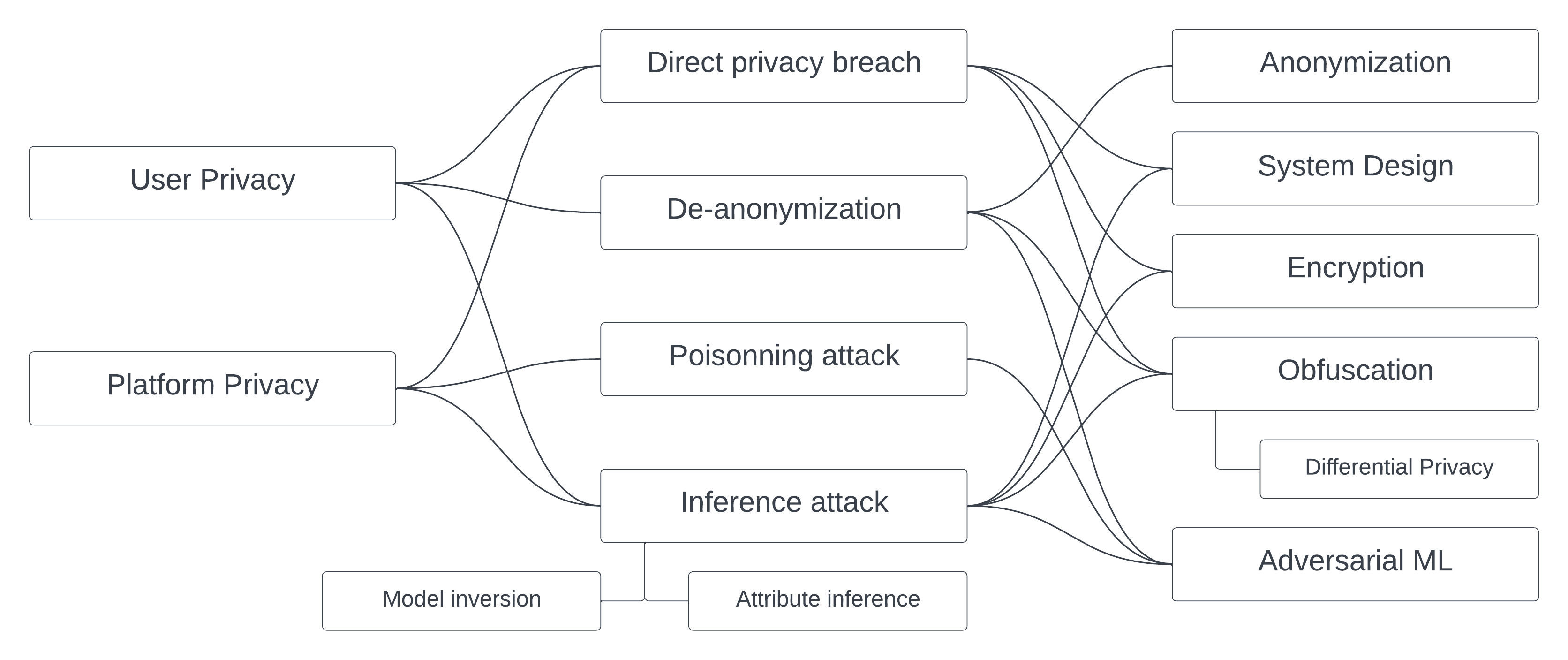}
    \vspace{-5pt}
    \caption{Ownership types (left), privacy threats (middle), and protection techniques (right).}
    \vspace{-10pt}
    \label{fig: privacy_solution}
\end{figure}

\subsection{Privacy Threats in Recommendation}
\label{sec: privacy_threats}

In this section, we summarize existing research about privacy threats in recommender systems and categorize them with respect to each type of ownership category.
The taxonomies are provided in Figure \ref{fig: privacy_solution}.
We denote ``privacy breach'' as the efforts that aim to directly access private information through hacking or monitoring.
This type of attack is mainly based on prior knowledge of the software implementation details of the communication interface (i.e. the APIs provided by the RS), which is out of the scope of this survey.
We illustrate the details of each of the remaining topics as follows.

\subsubsection*{\textbf{De-anonymization}}

Recommender system may disclose its data to third parties or to the public, e.g. e-commerce services open their consumer log data to researchers.
Since the data in recommender system typically includes comprehensive user information, the system has to protect users' Personally Identifiable Information (PII) in the data through anonymization, protecting user information \cite{krishnamurthy2009pii, ohm2009broken}.
Such information includes but is not limited to the user ID, user interaction history, and user profile features such as gender, age, and address.
In this context, the PII is user's private information and the full ownership should belong to the user.
However, studies have shown that one can still de-anonymize (or re-identify) user identities by 1) cross-linking them with external information \cite{ganta2008composition, narayanan2008robust, li2017privacy}, or 2) inference from the partial observation of the same sample as we will explain later.
For simplicity, we denote de-anonymization as the first scenario which requires attribute linking with external data, while the latter is denoted as inference attack.

\subsubsection*{\textbf{Inference attack}}

Users or the recommender system platform may deny the upload or sharing of critical information to physically maintain privacy. 
However, studies have shown that it is still possible to accurately infer the protected attributes merely by using public information.
In terms of the user privacy, works have shown that the user's private information can be inferred by their behavior history \cite{bhagat2007applying,calandrino2011you,gong2016you}, rating data \cite{weinsberg2012blurme}, user interests \cite{chaabane2012you}, social relations \cite{he2006inferring,dey2012estimating}, or even the writing style of the textual feature \cite{rao2010classifying}.
This is also referred to as the attribute inference attack, and the attacker could be either the RS or external entities that have obtained public information about the user and the RS.
In terms of the platform privacy, the existence of the recommendation also opens the door to the model inversion attack \cite{fredrikson2015model}, where attackers can observe the behavior of the model and then reconstruct the private information.  
For example, by sending queries and receiving feedback from the RS, one can reconstruct the missing private attributes of a sample \cite{hidano2017model,mehnaz2022your}, infer if a certain record (or user) is present in the training data of the model \cite{shokri2017membership,zhang2021membership} (a.k.a. membership inference attack), and even recover the model parameters \cite{tramer2016stealing}.
In general, the inference attack does not always require auxiliary data with correspondence of attributes like that in de-anonymization but always relies on the data correlation or the predictive model to engage the inference.

{
\subsubsection*{\textbf{Poisoning attack}}

This type of threat aims to indirectly control the recommendation model (regarded as private information) by modifying the training data through the provided trusted channel.
Compared with the previously mentioned attack methods that ``read'' the private information, the poisoning attack distinguishes itself for it aims to ``write'' the private information.
In general, the poisoning attack can be regarded as an optimization problem and it is also related to the problem of RS robustness as we will show in Section \ref{sec: robustness}.
}



\subsection{Privacy Protection Techniques}\label{sec: privacy_protection}
\subsubsection*{\textbf{Anonymization}}

The confrontation between anonymization and de-anonymization has been around for years, and the goal is to hide certain user details (i.e., PII) when publishing the dataset to third parties.
There are several studies proposed to solve this privacy issue based on the estimated risk of identifying a user in a system, such metrics include $k$-anonymity \cite{sweeney2002k,chen2012privacy}, $l$-diversity \cite{machanavajjhala2007diversity}, and $t$-closeness \cite{li2007t}.
Another effective solution is to replace the detail with high-level group information (i.e., through clustering) \cite{manolis2012disassociation}.
As described in the previous section, simply replacing attributes with pseudonyms does not guarantee privacy since one can still de-anonymize the private information even with a little bit of auxiliary background knowledge \cite{ganta2008composition, narayanan2008robust, li2017privacy}.
Thus, perturbation-based approaches such as differential privacy are introduced to alleviate this problem as we will illustrate later in this section.

\subsubsection*{\textbf{System Design}}

This type of solution is one of the earliest works that discuss the system-level privacy issue. Some typical insights include 1) letting users know about the privacy risks and acquire consent, 2) certification and verification between multiple parties \cite{cissee2007agent}, 3) dynamic authentication and temporal limits for access, and 4) risk reduction by distributed data storage and computation.
The last solution with distributed data can be further categorized into the service-side distribution and the user-side distribution.
The service-side distribution involves multiple functional participants that jointly provide the recommendation service \cite{aimeur2008alambic}, and the user-side distribution (e.g. federated learning or block-chain) allows users to keep their own data on personal devices without uploading to the central server \cite{canny2002collaborative,jiang2019towards,chai2020secure,muhammad2020fedfast,wu2021fedgnn,palato2021federated}.

\subsubsection*{\textbf{Encryption}}

The RS communicates with user devices and third-party services, as a result, it introduces the risk of data leakage (i.e. through external monitoring or hacking) whenever there is message transfer.
Thus, encryption techniques are used to encode the transferred information so that no one except for the owner can easily decipher the message \cite{canny2002collaborative}.
One widely adopted method is the homomorphic encryption \cite{zhan2010privacy,calandrino2011you,erkin2012generating,sobitha2016role,chai2020secure}, which allows third party computation without decryption or computational error.
Another method that provides secure computation is the garbled circuits with public-key encryption \cite{nikolaenko2013privacy}, where the collaborative filtering model is encrypted as a circuit, and multiple parties jointly optimize the model without revealing their input data.
In general, the encryption techniques are often used as fundamental building blocks for multi-party and communication-rich frameworks such as federated learning and secure multi-party computation \cite{bonawitz2017practical}.
Though these approaches provide accurate computation and guaranteed privacy, it also induces non-trivial extra computational cost \cite{badsha2016practical}.

\subsubsection*{\textbf{Obfuscation and Differential Privacy}}

When the leakage of the data (either through intentional publication or unintentional leakage) is inevitable, an alternative solution is adding noise to the data so that it can disguise the actual value.
This type of solution is effective in RS since the recommendation model can usually be expressed as an ``aggregation'' of individual information.
Intuitively, the introduced noise should significantly change the value of each record (or user) such that the actual value is hard to infer, and the noise should also be sufficiently small such that the aggregated value is still statistically accurate.
Early works studied simple obfuscation and perturbation methods \cite{polat2003privacy,weinsberg2012blurme}, and later approaches formulate this as differential privacy \cite{mcsherry2009differentially,zhu2014effective,berlioz2015applying,chen2021differentially}.
Compared to encryption methods, these noise-injection solutions are more efficient in practice but introduce additional computational error and utility loss.

\subsubsection*{\textbf{Adversarial Machine Learning}}

One critical problem of obfuscation-based methods is the loss of recommendation utility, as a result, some recent works propose to learn special noise distributions that achieve differential privacy without compromising the utility by formulating the noise finding task as a machine learning problem \cite{jia2018attriguard,chen2019adversarial}.
Note that this type of solution also belongs to the family of differential privacy methods and it is specifically designed for defending against user attribute inference attacks.
A different type of adversarial machine learning solution is training an adversarial model that mimics the attackers' behavior for a wider range of attacks \cite{he2018adversarial,beigi2020privacy,anelli2022adversarial}.
This adversarial model is later utilized to train an adversarial-aware recommendation model that is robust against such attacks without significant loss of recommendation utility.
Both types of adversarial learning are closely related to the system robustness, so it is particularly suitable for fighting against poisoning attacks and inference attacks.

\subsection{Relationship with other Trustworthy Recommendation Perspectives}
\label{sec: privacy_open_problems}
{
Privacy violations in recommender systems can have a significant impact on consumers and the platform itself, which can ultimately undermine the system's trustworthiness. 
Firstly, users may be negatively impacted by privacy violations in a recommender system in several ways. 
For instance, if the system collects and uses personal data without the user's acknowledge or consent, users may feel violated and lose trust in the system, which can lead to decreased engagement and revenue for the platform. 
Furthermore, users may choose not to use the system, resulting in limited exposure to certain items, and reduced ability to make informed decisions.
Secondly, privacy violations can negatively impact the platform itself. If the system is perceived as violating user privacy, it can result in reputational damage, decreased user trust, and engagement. Users may leave the platform, resulting in decreased revenue, and potential legal consequences.
Therefore, to ensure the trustworthiness of recommender systems, it is essential to design them with privacy in mind. 
By designing recommendation systems with privacy in mind, users can feel confident that their personal data is protected, which can lead to increased user trust, engagement, and revenue for the platform, as well as increased revenue for items being recommended. Also, privacy is closely related with other trustworthy perspectives in recommender systems:
}

\begin{itemize}
    
{
    \item \textbf{Explainable Privacy} aims to explain how the recommender system guarantees privacy, why it is effective against certain privacy risks, what are possible side effects, and why the suggested privacy setting is optimal for users. 
    This is because some existing privacy protection techniques are built based on deep learning models that are non-trivial to interpret. However, it is sometimes highly demanded to explain how the system guarantees privacy and why it is effective.
    This topic is one of the new research areas in explainable AI and it is closely related to the research of explainable security that recently gained research attention \cite{sarica2020explainable,meske2022explainable}.
    Explainable Privacy is also helpful for users' privacy setting decisions. For example, the system can recommend the optimal privacy setting for each user and meanwhile explain to users why the recommended setting is optimal \cite{mosca2021elvira,yu2018leveraging,sanchez2020recommendation}. The system may also explain to users why it needs access to certain information so as to provide the best service for users, which helps to gain trust from users compared with using such information without explaining to users \cite{kunkel2019let}.

    \item \textbf{Fairness-aware Privacy} aims to promote privacy protection in a fair manner. This can be interpreted on two dimensions: 1) The system's privacy scheme should provide fair and nondiscriminatory privacy protection to all users irrespective of their group membership or level of vulnerability \cite{ekstrand2018privacy}, and 2) The system should not lead to unfair service quality due to its privacy protection regime \cite{fernando2021missing,zhang2021assessing,noiret2023fairly}.
    Moreover, at the platform level, when multiple recommender systems collaborate and transfer their knowledge to boost each other's performances (as in cross-domain recommendation), protecting the privacy of one participant may restrict its contribution to the overall collaboration, which would be unfair to systems that offer more information. In general, an unfair privacy protection policy may become even more challenged when participants wants to protect more while contribute little, forming a multi-party min-max game and an agreement is hard to form. As a result, fairness should be taken into account when developing privacy protection policies \cite{resheff2018privacy,ekstrand2018privacy}.

    \item \textbf{Robust Privacy} aims to guarantee that the recommender system's privacy protection mechanism is resistant to attacks or changes in data distribution. Techniques that are based on theoretical encryption and anonymization have strong theoretical privacy guarantee. 
    Empirical methods such as differential privacy techniques and adversarial machine learning has weaker protection and are sometimes unstable, but they appear to be much more efficient \cite{li2021survey}. However, it is worthwhile to investigate the cases when these empirical methods fail in privacy protection. Additionally, privacy protection may systematically downgrade the accuracy of information transfer, introducing new robustness challenge to the recommender system. In certain systems such as decentralized recommendation, there may even exist inevitable trade-off between robustness and privacy and thus we have to consider the best balance between the two \cite{cheng2009trading}. For example, federated architecture helps to protect user privacy by keeping user's private data on the user's own device. However, on-device data could be easier to attack for malicious parties compared to on-cloud data which is protected by professional data scrutiny team. As a result, careful consideration should be taken for the robustness of privacy protection mechanisms.

    \item \textbf{Controllable Privacy} provides users with the ability to control the type and amount of personal information that can be used by the system in generating recommendations \cite{anelli2022user,kelley2008user}. This can be achieved through various methods, such as giving users the option to opt-out of data collection or to limit the types of data that are used. For example, users can control the type and amount of sensitive data that leaves their devices in a federated recommender system \cite{anelli2022user}. This allows users to explore the trade-off between privacy and service qualify and find out the sweet point by themselves. Controllable privacy has several potential benefits for users. First, it can help to address concerns around privacy and data protection, which have become increasingly important in today's digital age. By giving users greater control over their personal information, recommender systems can help to build trust and confidence in their services. Second, controllable privacy can help to improve the quality of recommendations by ensuring that only the most relevant and appropriate data is used in the recommendation process. This can help to reduce noise and improve the accuracy of recommendations, ultimately leading to a better user experience.
}

\item \textbf{Privacy Attacks on Explainability.}
    Privacy attacks may target the recommender system's explanation model, since the explanation model may reveal critical information about the underlying predictive model \cite{kuppa2020black}. As a result, it is important to explore privacy-preserving explanation models that explain the model decisions and meanwhile protect system security. Both topics are not yet adquently discussed in recommender system research.

\item \textbf{Privacy in Decentralized Systems.}
    Recent developments on Internet of Things (IoT) systems and blockchain have brought new ideas for further distributing the control and computation towards edge devices.
    Providing recommendation services in such systems can systematically reduce privacy risks because of the distributed data storage and model computation \cite{lisi2020practical,deng2020edge,gong2020edgerec}.
    However, it also indicates that the central service will have weak or even no control over the RS.
    Different from the centralized RS which only involves communications between a user's edge device and the central server, decentralized RS involves direct communication between edge devices, which introduces new privacy challenges yet to be explored.
\end{itemize}


%% file: sections/robust.tex
\section{Robustness}\label{sec: robustness}
Recommender systems enhance the efficiency of information seeking and benefit both customers and producers. However, they can also expose users to threats in terms of robustness, leaving room for third parties to manipulate recommendation outcomes through profile injection attacks, also known as shilling attacks \cite{si2020shilling,deldjoo2021survey}. These attacks are often motivated by malicious intent, such as personal gain of illegitimate profits, market penetration of certain items/brands, or even causing malfunction of the system.
Given that recommender systems are now used in many high-stake decision-making scenarios, this vulnerability raises concerns about how machine learning techniques can be safely adopted in recommender systems and how these systems can be carefully designed to be robust and trustworthy against aggressive provocation from attackers \cite{mobasher2007toward,anelli2022adversarial}. It is crucial to address these issues to ensure the continued reliability and utility of recommender systems.

\subsection{Taxonomy of Attacks in Recommendation}
{
Adversarial attacks refer to the deliberate introduction of malicious perturbations to recommendation data or models with the goal of manipulating recommendations or exploiting vulnerabilities in the system \cite{mobasher2007toward,si2020shilling}.
Adversaries with malicious intent can introduce adversarial perturbations through various means and under different scenarios. Commonly, such data perturbations are implemented by adding fictitious user profiles to user-item interactions, modifying user attribute information (e.g., age, gender, occupation), or altering item-side information, such as movie attributes and descriptions. These perturbations can be introduced with the objective of manipulating the underlying recommendation model and exploiting vulnerabilities in the system. 
Technically, adversarial attacks against recommendation models can be categorized according to various dimensions \cite{anelli2022adversarial,deldjoo2021survey}, such as the timing of the attack, the intent of the attackers, the size of the attack, and the knowledge of the attackers, which are briefly introduced as follows.
}

\begin{itemize}
    \item \textbf{Attack Timing}. Based on when the attacks occur in the learning pipeline, they can be categorized as \textit{poisoning attacks} and \textit{evasive attacks}. On one hand, poisoning attack happens before the model is trained and the attackers add attacking data points into the training data, causing the trained model to make erroneous predictions. As a result, poisoning attacks can directly influence the trained model. On the other hand, evasive attacks happen after the model is already trained and thus do not influence the trained model itself, instead, it aims to inject fake results into the model output while avoiding being detected by the model \cite{anelli2022adversarial,biggio2013evasion}.
    
    \item \textbf{Attackers' Intent}. Different attackers may have different intents. ``Push attack'' and ``nuke attack'' are two basic intents, where an attacker might insert fake profiles into RS to make an item/producer more likely to be recommended (``push'') or less likely (``nuke'') by the recommendation algorithm \cite{anelli2022adversarial}. Apart from these, the attacker may even attack the system just to make it recommend irrelevant items to users so as to lower users' trust, which is also known as ``untarget attacks''.
    
    \item \textbf{Attack Size}.The size of an attack can be evaluated through various means, with the most commonly used metric being the number or percentage of profiles that an attacker injects into a system \cite{williams2007defending}. Empirical studies suggest that in many commercial recommender systems, the total number of injected profiles typically ranges from 1\% to 15\% due to the considerable effort and information required to carry out such an attack \cite{si2020shilling}.

    \item \textbf{Attackers' Knowledge}. It refers to the attackers' knowledge of the data distribution in the recommendation system's database. A high-knowledge attack is one that requires very thorough knowledge of the data distribution in a recommender system's database, and if the attackers reproduce the precise details of the data distribution within the profile database, then this attack is called a perfect-knowledge attack \cite{williams2006profile}. Furthermore, a low-knowledge attack is one that only requires system-independent knowledge such as that obtained by consulting public information sources \cite{anelli2022adversarial,mobasher2007toward}.
\end{itemize}

{
\subsection{Attack Methods}
Based on existing literature, there are two main classes of attacks on RS: 1) hand-engineered shilling attacks, and 2) machine-learned adversarial attacks. 
The former relies on hand-engineered fake user profiles (typically a rating profile) injected into the system, while the latter is machine-learned attack optimized to find minimal perturbation of the user-item rating matrix or user/item content data to influence the recommendation performance. Moreover, machine-learned attacks can also be seen as a novel type of shilling attack that applies the adversarial learning paradigm for generating poisoning input data.

\subsubsection*{\textbf{Hand-engineered Shilling Attack}}
Shilling attacks against recommender systems have established literature. Since the early 2000s, the literature was focusing on building hand-crafted fake profiles whose rating assignment follows different strategies, such as random \cite{10.1145/988672.988726}, popular \cite{o2004evaluation}, love-hate \cite{mobasher2007toward}, bandwagon \cite{o2005recommender}, and average \cite{10.1145/988672.988726}, among others.
Specifically, given a user-item interaction matrix, the goal of a shilling attack is to insert fake profiles into the matrix in order to affect the predicted ratings and/or diminish the performance of the system to reveal the attackers' engineered and illegitimate targets, e.g., pushing some targeted items into the top-$K$ recommendation list to improve their market penetration \cite{gunes2014shilling}. 

However, there are several problems associated with hand-engineered shilling attacks. {First, executing these attacks requires significant effort and resources, encompassing tasks like creating multiple fake accounts and generating seemingly authentic user behavior, which limits their scalability and practicality in real-world situations. Second, hand-engineered attacks are often targeted and can only affect a small subset of users or items, limiting their capacity to cause widespread or systemic disruptions in recommendation systems.} Third, hand-engineered attacks can be easily mitigated by implementing countermeasures such as outlier detection, which makes them less effective against modern recommendation systems.

\subsubsection*{\textbf{Machine-learned Adversarial Attacks}}
The research works focusing on machine-learned adversarial attacks against RS have recently received considerable attention from the research community \cite{cao2020adversarial, liu2020certifiable}.
Based on their level of granularity, researchers mainly classify machine-learned adversarial attacks into three categories: 1) adversarial perturbation of model parameters, 2) adversarial perturbation of contents, and 3) machine-learned data poisoning attacks \cite{deldjoo2021survey,anelli2022adversarial}.

\begin{itemize}
    \item \textbf{Adversarial perturbation of model parameters}.
    The primary goal of this type of attack is to add adversarial perturbation on the user and item representations in the latent space.
    For example, \citeauthor{he2018adversarial} \cite{he2018adversarial} study adversarial attack strategies against the recommender system's parameters with a specific focus on the robustness of BPR-MF \cite{bpr} against adversarial perturbations on the user and item embeddings.
    Specifically, the authors first generate the perturbations based on the Fast Gradient Sign Method (FGSM) proposed by \citeauthor{goodfellow2014explaining} \cite{goodfellow2014explaining}, then add this perturbation to the model parameters, and finally generate the recommendation list with this perturbed model parameters.
    After exposing the parameters of BPR to both adversarial and random perturbations, they claim that the value of Normalized Discounted Cumulative Gain (NDCG) is decreased by -21.2\% and -1.6\%, respectively, which shows a huge difference.
    Several other recent works have performed similar FGSM perturbation against different recommender systems, such as visual-based recommender \cite{tang2019adversarial}, factorization machines \cite{chen2019adversarial}, deep neural network models \cite{yuan2019adversarial, tan2023towards} and auto-encoder models \cite{yuan2018adversarial}.
    The research summarized above indicates that model-based recommenders are highly susceptible to limited adversarial perturbations applied to model parameters, which can completely misuse a recommender's utility by slightly perturbing their latent factors. Moreover, even a small perturbation can result in a significant loss of accuracy, which could occur in a real scenario with only a few malevolent users whose actions cause a model update that has a significant negative impact on performance \cite{he2018adversarial}.

    \item \textbf{Adversarial perturbation on content data}.
    Adversarial attacks that manipulate the content data of users and items are a growing concern, especially for content-based and hybrid recommender systems. In a notable example, DeepWordBug, a black-box attack strategy proposed by \citeauthor{gao2018black} \cite{gao2018black}, generates a spam movie review that deceives a deep RNN-based classifier into incorrectly classifying it as a positive message. This type of attack can significantly alter the recommendations of review-based recommenders by distorting the actual feedback from users. Other types of recommender systems, such as social-based and sequence-aware, are also vulnerable to similar attacks \cite{wei2020heuristic}. 
    These examples illustrate how any hybrid or content-based recommender system may be vulnerable to adversarial attacks on their input data. As such, developing defenses against such attacks is becoming increasingly important for the development of trustworthy recommender systems.

    \item \textbf{Machine-learned data poisoning attacks}.
    In recent years, machine-learning-based data poisoning attacks have gained attention due to their ability to optimize attack goals automatically. 
    Similar to previously mentioned shilling attacks, this class of attacks targets the learning algorithms by manipulating the data used for training these models, which happens before the recommendation model is trained. However, the difference is that they are learned through specific optimization procedures maximizing the adversary’s goal automatically.  
    For example, \citeauthor{fang2020influence} propose a data poisoning optimization method for top-$N$ recommendation systems \cite{fang2020influence}. In this work, the attacker's objective is to promote a particular item to as many normal users as possible and maximize the hit ratio, which is defined as the fraction of normal users whose top-$N$ recommendation lists include the target item. To achieve this goal, the authors propose a gradient-based optimization algorithm based on the influence function approach, which claims that the top-$N$ recommendations are mainly affected by a subset $S$ of influential users. 
    Finally, they propose a gradient-based optimization algorithm based on $S$, named $S$-attack, to determine the fake users’ rating scores, and add the fake users to the recommender system.
    Another line of research utilizes reinforcement learning (RL) to generate fake user profiles and launch black-box attacks. One such framework is PoisonRec, which employs a model-free reinforcement learning approach for this purpose \cite{song2020poisonrec}. The framework constructs a Biased Complete Binary Tree (BCBT) to model the item sampling process, which significantly reduces the time complexity in a hierarchical action space. By generating fake user profiles, PoisonRec can manipulate the feedback received by the recommender system, leading to biased recommendations. Recently, generative positing attack has also been used to attack graph neural network based recommender systems \cite{nguyen2023poisoning}.
    This highlights the importance of developing countermeasures to detect and prevent such attacks, in order to enhance the trustworthiness of recommender systems.
    
\end{itemize}

\subsection{Defence Methods}
Modern recommendation systems are highly vulnerable to various attacking methods, which motivates researchers to design countermeasure strategies against adversarial attacks. In this subsection, we introduce representative defense methods that improve the robustness of recommendation systems. There are generally two pathways to defend against an adversarial attack in recommendation systems: (1) Detection methods to localize anomalies, such as fake user profiles, and (2) Adversarial Robust Training to enhance the robustness of recommender systems against adversarial attacks.
Specifically, the detection methods aim to identify and localize anomalies by analyzing and comparing the behavior patterns of different users or items. Meanwhile, adversarial robust training involves augmenting the training data with adversarial examples that are generated by perturbing the original data with small and imperceptible perturbations. This process enhances the robustness of the recommender system by exposing it to potential adversarial attacks during training and thus, improving its ability to withstand such attacks during testing.

\subsubsection*{\textbf{Detection methods}}
The detection of shilling attacks has been the subject of much research, with a considerable number of works having been developed \cite{yang2016re,bhaumik2011clustering,zhang2015catch,gunes2014shilling,si2020shilling}. These methods can be broadly classified into supervised classification methods and unsupervised clustering methods. Supervised classification methods typically involve feature engineering followed by algorithm development. For example, \citeauthor{yang2016re} \cite{yang2016re} use three well-designed features extracted from user profiles, namely the filler size with maximum, minimum, and average ratings, where filler size is defined as the ratio between the number of items rated by the user and the number of items in the entire recommender system. These features are then analyzed with statistical tests and classified using a variant of AdaBoost called re-scale AdaBoost (RAdaBoost) for attack profile detection. In contrast, unsupervised clustering methods aim to group individuals into clusters and then identify and eliminate suspicious ones. For instance, \citeauthor{bhaumik2011clustering} \cite{bhaumik2011clustering} use $k$-means clustering to detect relatively small clusters as attack groups of user profiles, while \citeauthor{zhang2015catch} \cite{zhang2015catch} develop a propagation method based on matrix iteration to calculate the likelihood of each user to be an attacked user based on a small seed group of known good users.


\subsubsection*{\textbf{Adversarial robust training}}
Adversarial robust training aims to develop attack-resistant recommendation algorithms, which is trying to reduce the influence of shilling attacks. Several recent studies have proposed methods for improving the robustness and reliability of recommendation models against shilling attacks. For example, in \cite{zhang2017robust}, \citeauthor{zhang2017robust} propose a robust collaborative filtering method by incorporating non-negative matrix factorization (NMF) with R1-norm, while in \cite{yu2017novel}, the authors' design a robust matrix factorization model based on kernel mapping and kernel distance. 
Moreover, to protect recommenders against Fast Gradient Sign Method (FGSM) attack, \citet{he2018adversarial} make the model robust to parameter perturbations by modifying the classical loss function of a recommender model (i.e., BPR-MF) by adding an adversarial regularization term. 
The application of adversarial training has been shown to improve the model's robustness against FGSM attacks, and then this training process has been applied to a variety of recommendation models, such as BPR-MF \cite{he2018adversarial} and VBPR \cite{tang2019adversarial}.
\citet{chen2019adversarial} also address the issue of robustness in tensor-based recommendations by incorporating adversarial training. Their approach aims to enhance the robustness of pairwise interaction tensor factorization \cite{rendle2010pairwise} for context-aware recommendations. By introducing adversarial perturbations during training, the model can learn to better distinguish between real and fake data, which can improve the overall robustness and reliability of the recommendations provided.
Moreover, \citeauthor{hidano2020recommender} \cite{hidano2020recommender} propose a defensive strategy leveraging trim learning, which exploits the statistical difference between normal and fake users/items, to make matrix factorization resistant to data poisoning. 
These approaches highlight the growing interest in incorporating adversarial techniques to improve the robustness and reliability of recommendation systems. By introducing adversarial perturbations and adversarial training during the model training process, these methods can increase the system's resilience against potential attacks, improving the overall quality of the recommendations provided.
}

\subsection{Relationship with other Trustworthy Recommendation Perspectives}
{
The reliability and robustness of a recommendation system are crucial to providing accurate and relevant recommendations. 
A lack of robustness can expose the system to various attacks and manipulations, which can severely compromise the integrity and quality of the recommendations. 
Consumers can be negatively impacted by the lack of robustness in a recommendation system in several ways. For instance, inaccurate recommendations may result in users being shown items that are not relevant to their needs or preferences, leading to frustration and decreased satisfaction with the system. Consequently, users may abandon the system, resulting in decreased engagement and revenue for the platform.
Lack of robustness can also affect items by reducing their exposure to potential customers. For example, if the system fails to recommend an item to a user who would have been interested in purchasing it, the item's sales may suffer, leading to decreased revenue for the item and the platform overall.
Furthermore, the lack of robustness can negatively impact the platform itself by reducing its perceived trustworthiness. Users may become skeptical of the system's recommendations if they frequently receive irrelevant or inaccurate recommendations. This can lead to decreased user trust, which can ultimately result in decreased engagement and revenue for the platform.
By designing recommendation systems with robustness in mind, users can be confident that the recommendations they receive are accurate, relevant, and trustworthy. Besides, robustness is also closed related to other trustworthy perspectives in recommender systems:
}

\begin{itemize}

{
    \item \textbf{Explainable Robustness} aims to provide explanations to users or system designers about why the system is resistant to attacks, or why certain component of the system is vulnerable to attacks and thus needs careful attention or improvement \cite{demontis2019adversarial}. The explainability of recommender systems plays a vital role in identifying potential vulnerabilities that can be exploited by adversaries. The methods developed to explore RS explainability can be used to implement both adversarial attacks and defenses on RS. By using explanation methods, the behavior of RS can be revealed, identifying why a specific recommendation is made concerning the current samples. These explanations provide insights that support the design of attack and defense algorithms. One example of the use of explanation methods for attacks on RS is explainability-based backdoor attacks \cite{xu2021explainability}. In this type of attack, an explainer for RS can be employed to identify the importance of items and guide the selection of the injected backdoor trigger position. The explainer helps to identify the key features or items that influence the recommendation decision and use this information to manipulate the system's behavior to the attacker's advantage. Moreover, defenders can use explainers for RS to identify potential locations of malicious item perturbations, which helps system designers to develop better defense strategies. For example, counterfactual explanation can help system designers to understand how the system would respond to possible data poisoning or profile injection attacks \cite{pawelczyk2022exploring} so that system designers can develop preventive strategies in advance. By understanding the system's behavior and potential attack strategies, defenses can be developed to detect and prevent such attacks, increasing the overall security of the RS.    
}

    \item \textbf{Fairness-aware Robustness} aims to guarantee that different users or user groups are fairly protected from attacks or fairly influenced in face of possible attacks \cite{ovaisi2022rgrecsys}. The research direction examines the fairness and vulnerabilities of recommender systems by exploring potential attacks that could cause certain items or producers to be recommended more or less likely. For example, users with fewer historical interaction records may experience more significant drop in recommendation quality when the system is under profile injection attack, since their recommendation can be easier to influence by injecting profiles \cite{gunes2014shilling}. As a result, it is important to explore how users can be fairly protected under attacks. Furthermore, existing works on machine learning (ML) have shown that adversarial robustness and fairness could negatively affect each other during training \cite{10.5555/3524938.3525692, xu2021robust}. It is possible that similar trade-offs between adversarial robustness and fairness may also exist in recommender systems, which is an area remaining largely unexplored.

    \item \textbf{Privacy-aware Robustness} aims to protect user privacy when possible attacks happen to the recommender system. Though most attacks to recommender systems aim to manipulate the recommendation results and promote target items in the recommendation list, some attacks aim at stealing users' private information from the system such as membership inference attacks \cite{zhang2021membership,wang2022debiasing,yuan2023interaction} or attribute inference attacks \cite{zhang2022comprehensive,zhang2021graph,zhang2020practical,chai2022efficient,aktukmak2019quick}. Besides, certain attacks may exfiltrate users' private data out of the system as a byproduct. For example, when the recommender system provides neighbour-user-based explanations for the injected profiles, the attacker may be able to infer the real users' information from the system \cite{coba2022recoxplainer,sharma2013social}. As a result, robustness research should also make sure users' information is protected from possible attacks \cite{lam2006you}.

    \item \textbf{Controllable Robustness} is essential in recommender system design. The concept of ``control'' in recommender systems should primarily be a tool for system designers, not users, to prevent adversarial user behaviors. It allows the system designers to generate controllable attacking examples or malicious activities to detect the weaknesses of the system and to improve the robustness of the system \cite{pereira2023the,zhang2019understanding}.
    

\end{itemize}

%% file: sections/controllability.tex
\section{Controllability}\label{sec: controllability}


{Controllability of AI is an important research problem \cite{yampolskiy2020controllability}, which is essential when users interact with intelligent systems and has been studied in the human-computer interaction (HCI) community for several decades \cite{he2016interactive, amershi2019guidelines, tsai2019controllability}.} In recommender systems, which actively interact with humans \cite{pmlr-v84-schmit18a, mcnee2006making, knijnenburg2011each, sulikowski2018human, shin2020users}, the importance of controllability can not be neglected. However, despite the recent successful improvement of the recommendation performance, the controllability issue in recommender systems has become a new major concern: most of the current RS are mostly uncontrollable by the system user and users can only passively receive the recommendation results. 
More specifically, when using a non-controllable recommender system, users can only passively choose to accept or not accept the recommendation results, but they can hardly control what the recommendation results they receive, and more importantly, what the recommender systems learn about them. In fact, controllability is an important aspect for building trustworthy recommender systems. Recent studies have shown that users may not be satisfied even with a high recommendation accuracy \cite{herlocker2004evaluating, mcnee2006being}, and increasing the users' controllability over recommender systems can increase the users' satisfaction and trust in the recommendation results {\cite{jugovac2017interacting, hijikata2012relation, knijnenburg2012inspectability, mcnee2003interfaces, xiao2007commerce, jannach2019explanations,zhang2020explainable, he2016interactive, verbert2013visualizing, swearingen2001beyond, parra2015user, rahdari2021connecting}}.

Recommender systems provide personalized recommendations based on user preferences, which are learned from users' interaction history. No matter what techniques are used for recommendation, the user preference construction process is crucial for making personalized recommendations \cite{atas2021towards}. However, traditional non-controllable RS fails to solve two problems: 1) the constructed user preferences may not be accurate, and 2) even with accurate user preferences, the users may not be willing to receive recommendations based on all of their preferences and information.
User controllable recommendation tackles the above problems by allowing users to manually express their preferences and intervene in the preference construction process via certain types of interactions. { Under the essential idea of collecting users' feedback to intervene in user preference construction for controllable recommendation, in this section, we broadly categorize existing research into two categories based on whether users directly change their preferences (explicit controllability) or indirectly change their preferences (implicit controllability). Explicit controllability involves systems where users explicitly acknowledge how they are changing their preferences in response to certain aspects, such as the types of music they like, the food they cannot eat, or the articles that they are particularly interested in. In contrast, implicit controllability does not involve direct interaction with user preferences. Instead, users intervene the learned preferences in an implicit manner, such as by clicking ``dislike'' on a recommended item. The system then learns how to update user preferences based on this behavior. 
\citeauthor{jannach2017user} \cite{jannach2017user} presents a detailed survey on controllable recommendation, categorizing user controllability based on the different phases of recommendation where intervention occurs. The approaches at the elicitation phase mainly involve explicit controllable methods, with the exception of conversational recommendation. The approaches to recommendation results primarily involve implicit controllable methods, with the exception of choosing recommendation strategy, for which users explicitly provide their preference on pre-set recommendation algorithms.
In the following parts of the section, we will introduce both types of controllability methods and then we will discuss the relationship between controllability and other aspects of trustworthy recommendation. 




}


\subsection{Explicit Controllability}
{
Research on controllable recommender systems based on explicit controllability let the users explicitly edit or update the user preferences. These systems assume that the users are not only aware of their desires and favors, but are also able to accurately express them. In these designs, users are knowledgeable or at least partially knowledgeable about how their feedback affects the construction of user preference.

The most common approach is to let users set their profiles or reweigh the pre-defined aspects or features to directly intervene in the preference construction process. }
For instance, many works in HCI focus on providing users with controllability through interactive recommendation. \citeauthor{ben2002meta} \cite{ben2002meta} introduces the concept of a meta-recommendation system that enables users to specify their desired bias towards the recommendation process through interactive interfaces. \citeauthor{verbert2013visualizing} \cite{verbert2013visualizing} propose a framework to control the recommender systems through visualization techniques, while other works \cite{bakalov2013approach, kangasraasio2015improving} allow users to adjust their profiles for better recommendations. In music recommendation, \citeauthor{millecamp2018controlling} \cite{millecamp2018controlling} present the results of a user study comparing two different visualisation techniques, a radar chart and sliders, to allow users to control Spotify recommendations. \citeauthor{liang2021interactive} \cite{liang2021interactive} studies two different visualizations and mood control methods to explore their helpfulness for new music genre exploration. \citeauthor{jin2019contextplay} \cite{jin2019contextplay} study user controllable music recommendation by allowing users to specify the context information such as mood, weather, and
location. \citeauthor{he2016interactive} \cite{he2016interactive} provides a comprehensive investigation into enhancing personalized controllability through interactive visualization techniques. Moreover, \citeauthor{hijikata2006content} \cite{hijikata2006content} propose a content-based filtering system for music recommendation using a decision tree, which enables users to edit the learned profiles on the tree. Similarly, in \citeauthor{hijikata2012relation} \cite{hijikata2012relation}, users are allowed to control a music recommender system by selecting the categories of the preferred items or editing their profiles. \citeauthor{terveen2002specifying} \cite{terveen2002specifying} propose a graphical interface model in a music recommender system, where users can stretch the bars in a preference histogram to increase or decrease their intent on different specifications.  \citeauthor{knijnenburg2011each} \cite{knijnenburg2011each} propose an interaction strategy where users can explicitly assign weights to each of the item attributes for controllability. Later in \citeauthor{bostandjiev2012tasteweights} \cite{bostandjiev2012tasteweights}, a more complicated hybrid music recommendation system is developed, where users can move sliders associated with the weights assigned on various items or features to observe the changes on the recommendation list. 

Other works combine scrutability with profile setting \cite{wasinger2013scrutable, lamche2014interactive}. For academic recommendation, \citeauthor{rahdari2019user} \cite{rahdari2019user} and \citeauthor{tsai2019exploring} \cite{tsai2019exploring} propose user-controlled interfaces for recommending papers in which users can choose the weights on different relevance sources. 
For educational recommendation, \citeauthor{rahdari2021connecting} \cite{rahdari2021connecting} propose a controllable recommendation approach to connecting students with research advisors.
For news recommendation, \citeauthor{ingvaldsen2015user} \cite{ingvaldsen2015user} develop a user interface design of the
SmartMedia news recommender prototype which gives the users control over the news stream compositions.
In \citeauthor{wasinger2013scrutable} \cite{wasinger2013scrutable}, a scrutable recommender system is used to recommend meals at a particular restaurant, where users are notified about how the preference contributes to the meal's personalization score. In \citeauthor{bhargav2021controllable} \cite{bhargav2021controllable}, a deep generative model based on disentangled learning is developed to control the generated recommendations. In \citeauthor{lamche2014interactive} \cite{lamche2014interactive} and \citeauthor{balog2019transparent} \cite{balog2019transparent}, textual, visual or attribute-based explanations are provided for users in the recommender system. In these works, if users are not satisfied with the recommendation list, they are able to explicitly modify the user preferences accordingly.

\subsection{Implicit Controllability}

Controllable recommendation methods based on explicit controllability face several issues. For example, the users may not always be fully aware of their preferences. The users may also be concerned about the effect or consequences of their revealed preferences due to the lack of transparency of the recommendation models.
Therefore, more recent works provide implicit controllability to the users. The crucial idea of implicit controllability is that the users do not directly manipulate the user profiles or favored features. Instead, they indirectly fine-tune their preferences when dynamically interacting with the recommender systems. For example, \citeauthor{Harper2015putting} \cite{Harper2015putting} allow users to interact with the recommendation results, in which users can re-rank the generated recommendation lists to fit their expectations. After re-ranking, the system will automatically update the users' preferences by re-weighting on two attributes: popularity and age of the items. \citeauthor{schaffer2015hypothetical} \cite{schaffer2015hypothetical} let users interact with the information used by the system to construct their preferences. In their developed interface, users are able to add and delete their historical interactions with items or modify the rating on the past items to observe the changes in recommendation results. { In \citeauthor{jin2018effects} \cite{jin2018effects}, the effectiveness of different controls are investigated in a hybrid visual user interface for a commercial music recommender, which allows users to control not only by explicitly changing their profiles, but also by providing feedback on recommendations results and algorithm parameters to implicitly change the recommendations.}
\citeauthor{harambam2019designing} \cite{harambam2019designing} provide several quantitative user study of both explicit and implicit user controllability in news recommendation scenarios. \citeauthor{tan2023user} \cite{tan2023user} propose controlling the user preferences learned by the recommender system by allowing users to interact with counterfactually generated recommendation explanations.

\subsection{Relationship with other Trustworthy Recommendation Perspectives}
{
The success of a recommender system highly depends on users' trust and satisfaction, which can be greatly influenced by the controllability of the system. Non-controllability can lead to negative outcomes, including reduced user satisfaction, decreased trust in the system, and decreased engagement with the platform. Effect on Users: Users' perception of having no control over the recommendations they receive can result in negative outcomes. Users may receive recommendations that do not align with their interests or needs, leading to dissatisfaction with the system and potentially causing them to leave the platform. Furthermore, if users are unable to provide feedback or adjust their preferences, they may perceive the system as unresponsive to their needs, reducing their trust in the system's recommendations. Effect on Items: Non-controllability can also affect items in a recommender system. When users cannot provide feedback on items they purchase for themselves or recommend to others, it can limit the exposure and sales of certain items. Consequently, this can lead to decreased revenue for items and the platform as a whole. Effect on the Platform: The non-controllability of a recommender system can negatively impact the platform's perceived trustworthiness. If users feel they have no control over the system's recommendations, they may become skeptical of its effectiveness. As a result, user trust can be reduced, leading to decreased engagement and revenue for the platform. 

As a result, to enhance the controllability of a recommender system, platforms can provide users with more opportunities to provide feedback, adjust preferences, and control the frequency and type of recommendations they receive. By providing users with greater control over the system, the platform can enhance user satisfaction, trust, and engagement. Additionally, incorporating mechanisms for users to provide feedback on items can improve the exposure and sales of certain items, leading to increased revenue for the platform. Ultimately, the success of a recommender system depends on its ability to balance user needs for control and the platform's goals for generating revenue and enhancing user engagement.
}

As research on user controllable recommendation continues to explore the feasible control options for users, it improves the users' trust and satisfaction in the recommender systems. { However, there are still obstacles to overcome to create a more trustworthy controllable recommender system, particularly regarding the correlation between controllability and other aspects of trustworthy recommendation. In the following, we present the relationships as well as some important directions that require further exploration.}

\begin{itemize}

{
    \item \textbf{Explainable Controllability} aims to explain to users how and why their preferences or interactions will help them to steer the system towards certain directions \cite{tsai2021effects,ngo2020exploring}. While more and more interaction options are provided to the users to increase their controllability and satisfaction, controlling the AI systems without understanding the underlying mechanism may cause errors or even be harmful rather than beneficial \cite{barbosa2020you}. In most of the existing works on controllable recommender systems, no matter whether the controllability is provided explicitly or implicitly, users only acquire an interface to intervene in the recommender system without knowing the consequences: in explicit controllability, users can directly interact with their preference information, but they do not understand how the preference contributes to the personalized recommendation results; in implicit controllability, users change the recommendation results to the desired direction so as to implicitly update their preference. However, they are not knowledgeable of how their interactions will influence the way that the system record their preference. Explainable controllability helps to explain to users how their actions will lead to certain outcomes so that they know the potential consequences before really implementing the control. Some research such as \cite{wasinger2013scrutable, lamche2014interactive} explains to the users which parts of their preference contribute to the recommended items before letting users change their preference, which may be a good attempt toward explainable controllability.
    
    \item \textbf{Fairness-aware Controllability} aims to mitigate bias and unfairness of the recommender system through user controllability and to prevent the system from introducing new bias or unfairness issues when users control the system. Although significant efforts have been made to improve recommendation models' fairness, user controllability may disturb the current system's behavior and reintroduce fairness issues. For instance, users may control the system by consistently disliking unpopular items, which can hinder the recommendation of new and potentially diverse items. Furthermore, prolonged control of the system may result in the recommendation of certain types of items, or items only from certain providers, which can raise extreme fairness concerns. For example, users may be trapped in filter bubbles when interacting with the recommender system, which may further lead to unfairness issues due to users' frequent clicks of popular items. Therefore, it is important to prevent user control from causing unintended and unfair biases in the recommendation process. By taking fairness into consideration when designing controllability methods, it can help users to gain control and meanwhile help the system to improve fairness, which provide benefits to both sides \cite{wang2022user}.
    
    \item \textbf{Privacy-aware Controllability} centers around protecting the privacy of users or item providers while still providing them with control over the recommender system. Although offering controllability can enhance users' experience, it can also introduce additional privacy risks. For instance, improving the controllability of a recommender system may require the collection of more user information, thereby increasing the possibility of a privacy breach. Moreover, in systems where users manually set their preferences, modified recommendation results may inadvertently disclose their preference settings. Malicious programs can exploit such controllability to extract private information about users. Additionally, when the ``items'' being recommended are human users, such as in friend recommendation \cite{shu2017user} or advisor recommendation \cite{rahdari2021connecting}, adversary users can more easily probe private information from the system since users can manipulate and compare the recommendation results through various controllability options. For example, users may control the system to recommend friends of a particular gender, potentially revealing the private gender information of the recommended users. Therefore, it is vital to investigate controllable recommender systems that also provide privacy protection \cite{wang2019privacy, walter2018user}.

    \item \textbf{Robust Controllability} aims to guarantee that the controllability function of the recommender system is robust to possible attacks. Recommender systems with controllability face unique challenges in terms of possible attacks or disturbances. For example, adversarial parties may manipulate the system based on data injection so as to lure the users to click manipulated items when users control the system \cite{jambor2012using}. Besides, an attacking method may significantly alter the system behaviors via the controllability options by modifying the (injected) user histories \cite{zhang2020practical, yue2021black}. Furthermore, an attacker could promote a particular item or disrupt the system's behavior by slightly altering users' preferences through the controllability back channel. Finally, An attacker could potentially uncover the underlying recommendation mechanism by studying the controllable process, thereby increasing the system's vulnerability. As a result, recommender systems that offer additional controllability settings may be more vulnerable to carefully designed attacks, and thus the robustness issue of controllable recommendation should be resolved in the future to safeguard against potential attacks on such systems. To solve the problem, it is important to make sure that the controllability functions are robust so that users can safely control the system towards their expected directions \cite{noh2015auro}.
}

\end{itemize}


    

%% file: sections/conclusion.tex
\section{Conclusion, Challenges and Opportunities}\label{sec: conclusion}
This survey summarizes the current developments and trends in trustworthy recommender system research, with the goal of facilitating and advancing future research and implementation of trustworthy recommender systems. We begin this survey by defining trustworthiness for recommender systems and illustrating their characteristics by categorizing trustworthiness principles. Following that, we introduce and discuss the recent advances in trustworthy recommender systems in terms of explainability, fairness, privacy, controllability, and robustness. We describe the fundamental ideas for each component, provide a detailed overview of existing methods for each of them, and then suggest the prospective future research directions for these elements, especially from cross-aspect perspectives.
Overall, the research field of trustworthy recommender systems is important and booming with a diverse set of approaches and applications, meanwhile, making recommender systems responsible and worthy of our trust is one of the biggest challenges that our research community needs to combat. We hope the survey equips researchers interested in this area with sufficient background and knowledge to meet this challenge.

Additionally, we conclude by discussing some of the open challenges and potential opportunities in the future development of trustworthy recommender systems.

\subsection{The Cost of Trustworthiness in Recommendation}
The adoption of trustworthy recommender systems relies on a thorough understanding and acceptance of the key concepts in this area. However, the impact of different objectives of trustworthiness on model performance from an algorithmic perspective remains inadequately understood. For instance, adversarial robustness improves the model's generalizability and reduces overfitting, but tends to adversely affect overall accuracy. Similarly, explainable- and fairness-aware recommender systems may also result in a loss of accuracy \cite{ge2021towards,ge2022toward}. Therefore, a shift towards trust-driven recommender systems may cause short-term side-effects such as longer learning time, slower development, performance drop, and increased costs of time and efforts associated with building such systems.
Nevertheless, researchers and practitioners should prioritize the long-term benefits of gaining the trust of all stakeholders so as to promote sustained use and development of these recommender systems. It is, therefore, imperative for practitioners to recognize the importance of incorporating a trust-driven approach in their decision-making processes when building recommender systems.

\subsection{The Trade-offs within Trustworthy Recommendation Perspectives}
As we have shown in Table \ref{tab: relationship}, there are a wealth of connections and interrelations between different aspects of trustworthiness. However, research has shown that there are frictions or trade-offs between these aspects in some cases, which we review here.

One of the key challenges is in privacy-protected recommender systems, which especially need to protect users' sensitive attributes such as age, gender, race, and location and avoid revealing such information to any party. However, such sensitive attributes are important and usually highly needed by algorithm developers to develop fairness-aware models so as to improve fairness treatment of users. Such algorithm developers may include the developers within the company, contracted third-party developers, or even non-contracted independent developers. As a result, there may exist an intrinsic conflict between promoting privacy and promoting fairness, since promoting fairness may require access to certain sensitive features while promoting privacy refrains from sharing such features, which has been pointed out as a critical challenge for privacy-protected machine learning \cite{zhou2021towards}. In some cases, the attributes that require fairness control (e.g. gender) are exactly the sensitive information that needs privacy protection \cite{liu2022fairness}. Therefore, it is crucial to find a solution that can achieve fairness without compromising privacy, especially in cases where the attributes that require fairness control are also the sensitive information that needs privacy protection.

Apart from the above privacy-fairness trade-off, algorithmic friction exists between the various perspectives of trustworthiness. Studies have shown that adversarial robustness and fairness can negatively affect each other during model training \cite{10.5555/3524938.3525692, xu2021robust}. Techniques used to improve adversarial robustness, such as adding perturbations to input features, may inadvertently introduce bias, affecting the fairness of the model. Similarly, certain fairness constraints, such as demographic parity, may result in a model that is less robust to adversarial attacks, as it may be more susceptible to overfitting to certain features. Furthermore, studies on fairness and explainability have demonstrated trade-offs between these dimensions \cite{agarwal2021trade, jabbari2020empirical}.

These frictional effects indicate that trustworthiness cannot be achieved by optimizing a set of disjoint criteria. Instead, compatibility between multiple requirements needs to be carefully considered when integrating them into a single system. Recent studies provide useful insights into joint optimization and trade-offs between different aspects of trustworthiness \cite{xu2021robust, zhang2019theoretically}. Therefore, simply combining systems to improve each dimension of trustworthiness separately does not guarantee a more trustworthy and effective end result. Instead, a comprehensive approach that accounts for the intricate relationships and trade-offs between different dimensions of trustworthiness is necessary.

\subsection{Limitations in Current Evaluations of Trustworthiness}
Despite increasing interest and efforts in research, the quantification of various aspects of trustworthiness remains challenging. Aspects such as explainability and controllability of recommendation systems are highly subjective, and are seldom evaluated quantitatively. This poses a significant obstacle in accurately assessing and comparing different models. For instance, in existing research, user studies are commonly used to evaluate proposed controllability methods. However, there is no generally accepted quantitative evaluation metric that can evaluate various controllability methods in a consistent manner.
For another example, the correlation between the offline metrics of explainability and actual user comprehension and utilization of recommendation explanations remains unclear and warrants further research. To address this issue, it is necessary to develop a reliable and robust evaluation framework that can measure different aspects of trustworthiness.

Furthermore, it is imperative to establish a scientific basis for assessing trustworthiness by developing quantitative evaluation methods. This approach will enable us to conduct empirical studies and objectively compare different systems, facilitating significant progress towards building more trustworthy and reliable recommender systems. By moving beyond philosophical discussions and establishing a scientific framework for the evaluation of trustworthiness, we can make more solid advance towards the various perspectives of trustworthiness. 